\documentstyle[12pt,epsf]{article}

\newcommand{\bmat}{\left(\begin{array}}
\newcommand{\emat}{\end{array}\right)}
\def\NPB#1#2#3{Nucl. Phys. B{#1} (19#2) #3}
\def\PLB#1#2#3{Phys. Lett. B{#1} (19#2) #3}

\def\PRD#1#2#3{Phys. Rev. D{#1} (19#2) #3}

\def\MODA#1#2#3{Mod. Phys. Lett.  {#1} (19#2) #3}

\def\yzero{\smash{\hbox{$y\kern-4pt\raise1pt\hbox{${}^\circ$}$}}}

\def\-{\hphantom{-}}
\def\ov{\overline}
\def\s2{\frac{1}{\sqrt2}}

\def\beq{\begin{equation}}
\def\eeq{\end{equation}}
\def\beqa{\begin{eqnarray}}
\def\eeqa{\end{eqnarray}}
\def\tr{{\rm tr \,}}
\def\Tr{{\rm Tr \,}}
\def\diag{{\rm diag \,}}

\def\IF{\relax{\rm I\kern-.18em F}}
\def\II{\relax{\rm I\kern-.18em I}}
\def\IP{\relax{\rm I\kern-.18em P}}

\def\NN{{\cal N}}

\def\Dsl{\,\raise.15ex\hbox{/}\mkern-13.5mu D} 

\def\IC{\bf C}
\def\IZ{\bf Z}
\def\IT{\bf T}
\def\IG{\bf \Gamma}
\def\z2z2{$\IC^3/(\IZ_2\times\IZ_2)$}

\def\id{{\bf 1}}
\def\Dthree{${\ov {{\rm D}3}}$}
\def\Dfive{${\ov {{\rm D}5}}$}
\def\Dseven{${\ov {{\rm D}7}}$}
\def\Dnine{${\ov{\rm D9}}$}
\def\Dtsix{${\widehat{{\rm D}6}}$}
\def\Dtseven{${\widehat{\rm D7}}$}
\def\Dteight{${\widehat{\rm D8}}$}
\def\ts{\tilde s}
\def\tc{\tilde c}

\topmargin
-1.5cm
\textwidth
15.5cm
\textheight
24cm
\oddsidemargin
0.7cm
\evensidemargin
1.2cm

\begin{document}

\makeatletter
\@addtoreset{equation}{section}
\makeatother
\renewcommand{\theequation}{\thesection.\arabic{equation}}
\pagestyle{empty}
\rightline{CERN-TH/2000-269, FTUAM-00/21, IFT-UAM/CSIC-00-27}

\rightline{\tt hep-th/0009135}
\vspace{1cm}
\begin{center}
\Large{\bf Type IIB Orientifolds without Untwisted Tadpoles, and non-BPS
D-branes \\ [10mm]}
\normalsize{R.~Rabad\'an$^1$,  A.~M.~Uranga$^2$} \\
\normalsize{$^1$ {\em Dpto. F\'{\i}sica Te\'orica, Universidad Aut\'onoma
de Madrid \\ }}
\normalsize{\em 28049 Cantoblanco, Madrid, Spain}

\normalsize{$^2$ {\em Theory Division, CERN \\ }}
\normalsize{\em CH-1211 Geneva 23, Switzerland}

\vspace*{1cm}

\small{\bf Abstract} \\[7mm]
\end{center}

\begin{center} \begin{minipage}[h]{14.0cm}
{\small
We discuss the construction of six- and four-dimensional Type IIB
orientifolds with vanishing untwisted RR tadpoles, but generically
non-zero twisted RR tadpoles. Tadpole cancellation requires the
introduction of D-brane systems with zero untwisted RR charge, but
non-zero twisted RR charges. We construct explicit models containing
branes and antibranes at fixed points of the internal space, or non-BPS
branes partially wrapped on it. The models are non-supersymmetric, but are
absolutely stable against decay to supersymmetric vacua. For particular
values of the compactification radii tachyonic modes may develop,
triggering phase transitions between the different types of non-BPS
configurations of branes, which we study in detail in a particular
example. As an interesting spin-off, we show that the $\IT^6/\IZ_4$
orientifold without vector structure, previously considered inconsistent
due to uncancelable twisted tadpoles, can actually be made consistent 
by introducing a set of brane-antibrane pairs whose twisted charge cancels
the problematic tadpole.
}

\end{minipage}
\end{center}

\newpage
\setcounter{page}{1}
\pagestyle{plain}
\renewcommand{\thefootnote}{\arabic{footnote}}
\setcounter{footnote}{0}

\section{Introduction}

Type IIB orientifolds \cite{sag1, dlp, horava, sag2} (see \cite{gp,
gjdp, fourdim, zwart, afiv,kr2,kr3} for more recent references) allow the
construction of large
classes of supersymmetric string vacua, which have played a prominent role
e.g. in the study of string duality. Recently, and partially motivated by
the new developments in non-supersymmetric strings and non-supersymmetric
states in string theory (see \cite{revnonbps,gaberdiel} for a review) and
their relation to brane-antibrane configurations, it has become natural to
incorporate antibranes in the construction of non-supersymmetric type IIB
orientifolds \cite{sugimoto,ads,au} (see also \cite{aiq,aadds,abg} for
further developments), thus enlarging the class of models available
through this construction. Interestingly enough, such formal developments
have led to a breakthrough in the construction of phenomenologically
realistic orientifold models \cite{aiq,aiqu}.

Another interesting advantage in allowing the introduction of antibranes in
the orientifold construction is that they can be used to cancel RR
tadpoles in models for which no supersymmetry-preserving solution exists. 
A prototypical example is provided by models with untwisted tadpoles whose
cancellation requires a net negative D-brane charge. Six-dimensional
\cite{ads,au} and four-dimensional \cite{aadds} examples of this type are
easily obtained by considering a particular choice of the orientifold
projection on twisted sectors (sometimes referred to in the literature 
\cite{blpssw,polchinski,intri} as `with vector structure'), whose untwisted 
crosscap tadpole requires a positive net number of antibranes for 
consistency.

In this paper we study another class of type IIB orientifolds whose
crosscap tadpoles cannot be canceled by using just D-branes, i.e.
preserving supersymmetry. The models contain non-zero twisted RR tadpoles,
but zero untwisted RR tadpoles. To render them consistent one needs to
introduce systems of D-branes contributing to twisted tadpoles, but with
zero charge under the untwisted RR forms. Such brane systems are 
necessarily non-supersymmetric. Two simple examples are for instance sets
of branes and antibranes sitting at fixed points in the internal space
\cite{au} (see e.g. \cite{senkink, senvortex} for such configurations in
the context of non-BPS states in string theory), or non-BPS branes passing
through fixed points (the so-called truncated branes \cite{stefanski}).

Since no supersymmetric solution to the tadpole conditions exists, these
models are absolutely stable against decay to supersymmetric configurations. 
However, a given non-supersymmetric configuration of branes satisfying the
tadpole conditions may decay to a different consistent (and
non-supersymmetric as well) configuration as some of the geometric moduli
are varied. The space of geometries splits into different phases, at each
of which at least one (and possibly several) of these configurations of
branes is stable and tachyon-free. At their boundaries some configurations
become unstable and decay, leading to interesting phase transitions in
which the low-energy field theory changes drastically. This situation is
analogous to that studied in the literature on non-BPS states in string
theory \cite{senkink, senvortex, revnonbps, gaberdiel}, and has been
considered in the orientifold context in \cite{abg}.

In this paper we present several explicit six- and four-dimensional
orientifolds with the above features, where the tadpoles are cancelled
using brane-antibrane pairs at orbifold fixed points. Tadpoles could also
be cancelled by the introduction of a suitable set of non-BPS D-branes. In
a particular six-dimensional example, we construct explicitly the set of
possible configurations of branes cancelling the tadpoles, and discuss
their stability regimes as the compactification radii are varied, leading
to a picture of the phase diagram for this model.

In \cite{abg}, an asymmetric orientifold model without untwisted tadpoles
but with twisted tadpoles was constructed, and several of these issues
were addressed. Our purpose in this paper is to illustrate the construction 
of orientifolds without untwisted tadpoles in the context of completely
left-right symmetric constructions. We hope that, since they admit a
geometric interpretation, the class of models we consider will allow a
more intuitive and systematic study of this type of string vacua.

Our investigation leads to an interesting insight into an old problem in
the orientifold context. The $\IT^6/\IZ_4$ orientifold without vector
structure contains a non-zero twisted tadpole with an unusual volume
dependence, which cannot be cancelled by the D9- or D5$_3$-branes
\footnote{In our notation, D5$_i$-branes wrap the $i^{th}$ complex plane,
while D7$_i$-branes are transverse to it. We also denote antibranes with a
bar.} present in the model (required by the existence of untwisted
tadpoles) \cite{afiv}. Therefore it has been considered that the model
admits no solution to the tadpole cancellation conditions, and hence is
inconsistent. Our analysis shows that the latter claims are actually
incorrect. In particular, we show that it is possible to add configurations 
of branes which cancel the problematic twisted tadpole, while give zero
contribution to the untwisted tadpoles (specifically pairs of
D5$_1$-\Dfive$_1$ and/or D5$_2$-\Dfive$_2$ branes). In this way, we
achieve the construction of explicit consistent compact $\IT^6/\IZ_4$
orientifolds where all tadpoles are cancelled. The correct statement
about the $\IZ_4$ orientifold without vector structure is then that it
contains a twisted tadpole, whose cancellation requires its own new set
of branes contributing to the twisted tadpole, but not to the untwisted
tadpoles. Notice that our point is different from that made in
\cite{aadds,kr2}, where the $\IT^6/\IZ_4$ orientifold studied is actually
different from that in \cite{afiv}, since it corresponds to a projection
with vector structure, for which the problematic tadpole simply vanishes.

The $\IT^6/\IZ_4$ model (and other four-dimensional models with analogous
features, like the $\IZ_8$, $\IZ_8'$, $\IZ_{12}'$ \cite{afiv}) is similar
to the orientifolds without untwisted tadpoles mentioned above, in the
following sense. Both types of models contain twisted tadpoles whose
cancellation requires the introduction of branes for which no untwisted
tadpole exists. We hope this note clarifies that such models can be
consistently completed by appropriate sets of D-branes, and illustrates
the power of brane-antibrane systems (or other non-BPS configurations of
branes) in making problematic models consistent.

The paper is organized as follows. In sections 2, 3, and 4 we study the
construction of certain families of type IIB orientifolds without
untwisted tadpoles. The general pattern in their construction is the
following: We consider type IIB theory compactified on a toroidal
orbifold, $\IT^4/\IG$ or $\IT^6/\IG$ (even though a more general
starting point, like a general CY or a non-geometric CFT compactification,
is conceivable). We mod out the theory by an action $\Omega g$, where
$\Omega$ is world-sheet parity, and $g$ is a geometric action with
$g^2\in \Gamma$ (required by closure). Equivalently, we mod out type IIB
theory by the orientifold group $G_{orient}$, generated by $\Gamma$ and
$\Omega g$. An element $\Omega h$ in the orientifold group generates a
tadpole (via a crosscap diagram) at $h$-fixed points for a RR field in the
$h^2$-twisted sector. Hence, to obtain models without untwisted tadpoles,
we need $G_{orient}$ to contain no elements $\Omega h$ with an order two
$h$.

In Sections 2, 3 we consider a simple realization of this idea. We
consider a $\IZ_{2N}$ action $\alpha$ acting crystallographically on
$\IT^4$ or $\IT^6$, and consider type IIB theory on $\IT^6/\IZ_N$, with
$\IZ_N$ generated by $\theta=\alpha^2$. We further mod out this theory by
$\Omega \alpha$, hence the orientifold group has the structure
\beqa
G_{orient} & = & \{  1, \Omega\alpha, \alpha^2, \ldots,
\Omega\alpha^{2k-1}, \alpha^{2k},
\ldots \}
\eeqa
For odd $N$, $\alpha^N$ is of order two, and the orientifold group element
$\Omega\alpha^N$ leads to untwisted tadpoles. The resulting models are
therefore T-dual to familiar models studied in the literature. On the
other hand, for even $N$, $\Omega$ is never accompanied by an order two
action, and the resulting models contain no untwisted tadpoles, leading to
new type IIB orientifolds. For some models there is an additional
(accidental) cancellation of twisted tadpoles, resulting in a set of
models which do not require open string sectors for consistency, i.e. a
set of consistent unoriented closed string theories. The generic case,
however, corresponds to models with non-zero twisted tadpoles, which we
cancel through the introduction of brane-antibrane pairs at fixed points
in the internal space. We construct several six-dimensional (Section 2)  
and four-dimensional (Section 3) examples of this type.

In Section 4 we consider a further possibility to build new
six-dimensional orientifolds without untwisted tadpoles, based on a
different structure of the orientifold group. The models are obtained by
modding out $\IT^4/\IZ_N$, for even $N$, by the orientifold action
$\Omega\Pi$, with $\Pi:(z_1,z_2)\to (z_2,-z_1)$.

In Section 5 we discuss tadpole cancellation in the $\IT^6/\IZ_4$
orientifold without vector structure of \cite{afiv}, and provide a
consistent solution to the tadpole conditions (including the problematic
twisted tadpole)  using D5$_1$-\Dfive$_1$ pairs, in addition to the
usual D9- and D5$_3$-branes. 

In Section 6 we describe the stability regions of different brane
configurations for the particular six-dimensional orientifold of
$\IT^4/\IZ_2$ studied in Section~2, and obtain the phase diagram for this
model. This Section can be read independently of Sections 3,4 and 5.
Section~7 contains our final conclusions.

\section{Six-dimensional examples}

Before considering the more involved case of four-dimensional models, it
will be convenient to make the above ideas explicit in the six-dimensional
context. In this section we consider orientifolds of type IIB theory
compactified on $\IT^4/\IZ_N$. The orientifold action is $\Omega'\alpha$,
where $\Omega'=\Omega R$, and $R$ reflects all compact directions
\footnote{Choosing other orientifold actions, like $\Omega\alpha$, leads
merely to T-dual versions of our models.}. Also, $\alpha$ is a $\IZ_{2N}$
crystallographic action on $\IT^4$, and $\theta=\alpha^2$ generates the
orbifold group $\IZ_N$. Following the general arguments in the introduction, 
such orientifolds produce zero untwisted tadpoles when $N$ is even, so the
only crystallographically consistent $\alpha$ is the $\IZ_4$ twist
$\alpha:(z_1,z_2)\to(e^{2\pi i \frac 14} z_1, e^{-2\pi i\frac 14} z_2)$.

This type of models was studied in the non-compact context (i.e. 
orientifolds of $\IC^2/\IZ_N$, where the crystallographic constraint can
be dropped) in \cite{pu}, whose basic results can be used in the compact
context as well. One important difference is that in the non-compact case
the cancellation of untwisted charge may be violated, since the
corresponding flux may escape to infinity through the non-compact
transverse dimensions. Hence, models in \cite{pu} used D5-branes to cancel
twisted tadpoles, but did not include \Dfive-branes to compensate for
their untwisted charge. Our compact models will however necessarily
contain equal numbers of D5- and \Dfive-branes. A minor difference is that
\cite{pu} considered a T-dual version of these models, with orientifold
action $\Omega' R_3(-)^{F_L}$, in which the crosscap contributions to the
tadpoles are reduced by a factor of four with respect to our models here.

As discussed in \cite{pu} and following \cite{polchinski,intri}, it is
important to notice the existence of two orientifold projections for even
$N$, $N=2P$, which we refer to as {\bf A}- and {\bf B}-projections, which
differ in their action on order two twisted sectors, and which is
correlated \cite{polchinski} with the choice of vector structure on the
D-brane Chan-Paton bundle \cite{blpssw}. We discuss them in turn.

\subsection{Models with {\bf A}-type projection}

The first type of projection, which we denote as {\bf A}-type, chooses
left-right symmetric (antisymmetric) states in the NS-NS (RR) order two
twisted sector. The closed string twisted sector at a $\IC^2/\IZ_N$ point
leads to $P$ hyper- and $P-1$ tensor multiplets of $D=6$ $\NN=1$
supersymmetry. When open string sectors are required, this projection is
correlated with a choice of Chan-Paton matrices without vector structure
of the form (\ref{cpone}) for D5- and D9-branes (and antibranes), and
orientifold actions (\ref{cptwosym}) for D5-branes and (\ref{cptwoanti})
for D9-branes. 

Evaluation of the relevant amplitudes leads to the following crosscap
twisted tadpoles \cite{pu} at each $\IC^2/\IZ_N$ orientifold point 
\beqa
{\cal T}_{\theta^k} & = & + 32 \delta_{k,1\,{\rm mod}\, 2} \cos(\pi k/N)
\quad k=1,\ldots,N-1
\label{crosscap1}
\eeqa
In this and similar formulae, we normalize these twisted orientifold 
charges in units in which the D5- and D9-brane twisted disk tadpole are
$4\sin(\pi k/N) \Tr \gamma_{\theta^k,5}$ and $\Tr\gamma_{\theta^k,9}$,
respectively. However, we prefer not to introduce disk contributions for
the moment, since  in some cases open string sectors may not be required
for consistency.

For the case of interest in the compact setup, the $\Omega'\alpha$
orientifold of $\IT^4/\IZ_2$, the above closed string tadpoles vanish, and
the theory is consistent without the introduction of open string sectors.
In fact this model has already appeared in \cite{gjdp}. The untwisted
closed string sector contains the $D=6$ $\NN=1$ supergravity multiplet,
the dilaton tensor multiplet, and two hyper- and two tensor multiplets. In
the closed twisted sectors, the four $\alpha$-invariant $\IZ_2$ fixed
point give one hypermultiplet each, while the remaining twelve $\IZ_2$
fixed points (paired by $\Omega'\alpha$) give a total of six hyper- and
six tensor multiplets each. The spectrum, shown in Table~\ref{sixone}, is
free of gravitational anomalies.
\begin{table}[htb]
\small
\renewcommand{\arraystretch}{1.25}
\begin{center}
\begin{tabular}{|c||c|}
\hline
{\bf Untwisted} & $\NN=1$ SUGRA + 3 {\bf T} + 2 {\bf H} \\
\hline
{\bf Twisted} & 6 {\bf T} + 10 {\bf H} \\
\hline
\end{tabular} 
\caption{Spectrum of the $\Omega'\alpha$ orientifold of $\IT^4/\IZ_2$ with
{\bf A}-type projection.}
\label{sixone}
\end{center}
\end{table}
\subsection{Models with {\bf B}-type projection}

The second type of projection, which we denote as {\bf B}, chooses
left-right antisymmetric (symmetric) states in the NS-NS (RR) order two
twisted sector. The closed string twisted sector for a $\IC^2/\IZ_N$ 
point contains $P-1$ hyper- and $P$ tensor multiplets of $D=6$ $\NN=1$
supersymmetry. When open string sectors are present, this projection
corresponds to a choice of Chan-Paton matrices with vector structure, 
of the form (\ref{cpthree}), for D5-, D9-branes, and orientifold actions
(\ref{cpfoursym}) for D5-branes and (\ref{cpfouranti}) for D9-branes. The
closed string twisted tadpoles at each $\IC^2/\IZ_N$ orientifold point
are
\beqa
{\cal T}_{\theta^k} & = & - 32 \delta_{k,1\,{\rm mod}\, 2} \quad
k=1,\ldots, N-1
\label{crosscap2}
\eeqa
In contrast with the previous case, the case $N=2$ leads to non-zero
$\theta$-twisted tadpoles, which we may cancel by introducing e.g. D5- and
\Dfive-branes. 

Let us discuss the construction of the {\bf B}-type orientifold of
$\IT^4/\IZ_2$. The untwisted closed string sector is as in previous
subsection, giving the $D=6$, $\NN=1$ supergravity multiplet, the dilaton
tensor multiplet, and two hyper- and two tensor multiplets. In the closed
string twisted sector, instead, each $\alpha$-invariant $\IZ_2$ point give
one tensor multiplet. The remaining twelve $\IZ_2$ points still contribute
a total of six hyper- and six tensor multiplets. We achieve cancellation
of twisted RR tadpoles by introducing an equal number of branes and
antibranes, distributed among the four $\Omega'\alpha$-invariant fixed
points $P$, and cancelling the crosscap tadpole (\ref{crosscap2}) for
$k=1$
\beqa
4\, (\,\Tr \gamma_{\theta,5,P} - \Tr \gamma_{\theta,{\ov 5},P}\,)\, - 32=0
\eeqa
The condition can be satisfied in different ways. A simple possibility is
to locate one set of D5-branes with $\gamma_{\alpha^2,5}=\id_8$ at two of 
the fixed points of $\alpha$, and two \Dfive-branes with $\gamma_{\alpha^2, 
\bar{5}}= -\id_8$ at the remaining two. The remaining $\alpha^2$-fixed
points are not fixed under $\Omega'\alpha$, and generate no crosscap
tadpoles, and do not require any branes. The complete spectrum for this
model is provided in Table~\ref{sixtwo}. It is easy to see that all
irreducible six-dimensional gravitational and gauge anomalies \cite{erler}
cancel. The remaining factorized anomaly is cancelled by the generalized
Green-Schwarz mechanism \cite{sagnan}. 

Another simple choice of open string sector satisfying all constraints is
to locate, at each of the four $\alpha$-fixed points, a set of four
D5-branes and four \Dfive-branes, with
\beqa
\gamma_{\alpha^2,5}=\id_4 \quad ; \quad \gamma_{\alpha^2,{\ov 5}}=-\id_4
\eeqa
which leads to local cancellation of their untwisted RR charge. The
corresponding open string spectrum at each such point is
\beqa
 {\rm Gauge\; Bosons} & SO(4)_{D5}\times USp(4)_{\ov{D5}} \nonumber\\
 {\rm Fermion_+} & (6;1) + (1;5+1) + (4;4) \nonumber\\
 {\rm Cmplx. \; Scalars} & (4;4) 
\eeqa
and leads to a different anomaly-free model.
Notice that, despite the branes and antibranes sit at the same fixed
point, the tachyon in the $5{\ov 5}$, ${\ov 5}5$ groundstates is projected
out by the orbifold projection. This reflects the fact that the branes
and antibranes carry different twisted charges and cannot annihilate into
the vacuum.
\begin{table}[ht]
\small
\renewcommand{\arraystretch}{1.25}
\begin{center}
\begin{tabular}{|c|c||c|}
\hline
{\bf Closed} & Untwisted & $\NN=1$ SUGRA + 3 {\bf T} + 2{\bf H} \\
\hline
             & Twisted & 10 {\bf T} + 6 {\bf H} \\
\hline\hline
{\bf Open} & {\bf 55} & $\NN=1$ Vector of $SO(8)^2$ \\
\hline
           &  ${\bf{\bar 5}}{\bf{\bar 5}}$ & Gauge Bosons of $USp(8)^2$ \\
\hline
           & & Fermion$_+ \;\;$ $({\bf {27}+{\bf 1}}, {\bf 1}) + 
({\bf 1}, {\bf {27}+{\bf 1}})$ \\
\hline
\end{tabular} 
\caption{Spectrum of the $\Omega'\alpha$ orientifold of $\IT^4/\IZ_2$ with
{\bf B}-type projection.}
\label{sixtwo}
\end{center}
\end{table}

The above two solutions can actually be continuously connected, by
nucleating of brane-antibrane pairs in the bulk, distributing them
among the fixed points, and annihilating fractional branes (notice however
that this process is suppressed by energetic barriers, hence does not
alter the stability of the configuration). In fact, it is possible to
construct a large set of models connected in this way. The general
tachyon-free configuration is obtained by placing D5-branes and
\Dfive-branes at the points $P_r$, $r=1,\ldots, 4$, with
\beqa
\gamma_{\theta,5,P_r}=\id_{N_r} \quad ; \quad \gamma_{\theta,{\ov 5},P_r}
=-\id_{8-N_r}
\eeqa
Twisted tadpoles are cancelled by the above choice, and untwisted
disk tadpoles of branes and antibranes cancel each other if $\sum_{r=1}^4 
N_r=16$. Also, the tachyonic ground states in  $5{\bar 5}$ and ${\bar 5}5$
sector are projected out. This class of models contains as particular cases
the possibilities considered above. Concerning which configuration is
energetically favoured and provides the global minimum of the model
(keeping geometry fixed), the analysis of forces between branes and
antibranes at orbifold singularities in \cite{sunil} seems to suggest
the preferred configurations should not contains branes and antibranes at
the same fixed point.

The models we have constructed are absolutely stable against decay to
supersymmetric configurations, since no supersymmetric configuration of
branes can satisfy the tadpole cancellation conditions. The models,
however, can suffer diverse phase transitions as geometry is varied. In
Section 6 we discuss the phase diagram of this model for the distribution
of branes leading to the spectrum in Table~\ref{sixtwo}. The reader
interested only in this analysis is encouraged to skip Sections 3, 4 and
5.

\medskip

As mentioned above, there are no other compact six-dimensional examples
using the orientifold group structure mentioned above (in Section~4 we
consider further six-dimensional examples, based on a different
orientifold group). In the non-compact setup, however, the orientifold
group structure of \cite{pu} defines a set of orientifold planes (in the
sense that are fixed under an orientation-reversing action), preserving
half of the supersymmetries, and which have no charge under untwisted RR
fields, and twisted charges (\ref{crosscap1}) and (\ref{crosscap2}) for
the {\bf A}- and {\bf B}-type projections, respectively. Even though it
would be interesting to explore the properties of these `twisted
orientifold planes' in more detail, e.g. in the boundary state formalism,
we leave the discussion here and turn to the four-dimensional case.

\section{Four-dimensional examples}

It is straightforward to consider the construction of four-dimensional
models analogous to the six-dimensional examples above. Namely we consider
a crystallographic $\IZ_{2N}$ action $\alpha$, with a vector of twist
eigenvalues ${\tilde v}=\frac{1}{2N} (a_1,a_2,a_3)$. We center on models
preserving supersymmetry in the closed string sector, and so impose
\footnote{Considering $\sum_i a_i=N\;{\rm mod}\; N$ also leads to $\NN=1$
supersymmetric models. It is possible to show, however, that they are
equivalent to the ones we consider.} $\sum_{i} a_i= 0 \; ({\rm
mod} \, 2N)$. Without loss of generality we may take $a_1$, $a_2$ odd and
$a_3$ even. We consider type IIB theory on the orbifold \footnote{Extension 
to orbifolds  $\IT^6/(\IZ_N\times \IZ_M)$, not considered in the present
paper, is straightforward.} $\IT^6/\IZ_N$, where $\IZ_N$ is generated by
$\theta=\alpha^2$, with twist vector $v=2{\tilde v}=\frac 1N(a_1,a_2,a_3)$. 
Our models are obtained by modding out by the orientifold action $\Omega'
\alpha$, with $\Omega'=\Omega R_1 R_2 R_3 (-1)^{F_L}$, where $R_i:z_i\to
-z_i$, and $F_L$ is left-handed world-sheet fermion number. The D-branes
surviving the $\Omega'\alpha$ projection, and which we may use to cancel
the tadpoles are D3-branes (at a point in the internal space) and
D7$_i$-branes (transverse to the $i^{th}$ complex plane), and the
corresponding antibranes. Use of other orientifold projection, like
$\Omega\alpha$ would lead to T-dual version of our models.

\begin{table}[ht]
\small
\renewcommand{\arraystretch}{1.25}
\begin{center}
\begin{tabular}{|c||c||c|}
\hline
$\Gamma$  & $\Omega'\alpha$       &$\theta$ \\ 
\hline
$\IZ_2$   &${\tilde v}={1\over4}(1,1,-2)$ & $v={1\over2}(1,1,-2)$ \\
\hline
$\IZ_4$   &${\tilde v}={1\over8}(1,3,-4)$ & $v={1\over4}(1,3,-4)$  \\
\hline
$\IZ_4$   &${\tilde v}={1\over8}(1,-3,2)$ & $v={1\over4}(1,-3,2)$  \\
\hline
$\IZ_{6}$ &${\tilde v}={1\over12}(1,-5,4)$ & $v={1\over6}(1,-5,4)$ \\
\hline
$\IZ_{6}$ &${\tilde v}={1\over12}(1,5,-6)$ & $v={1\over6}(1,5,-6)$ \\
\hline 
\end{tabular}
\caption{Twist eigenvalues for the $\Omega'\alpha$ orientifolds of
$\IT^6/\IZ_N$ orbifolds.}
\label{ZN}
\end{center}
\end{table}

Crystallographic actions on $\IT^6$ compatible with unbroken
supersymmetry have been classified in the early orbifold literature
\cite{dhvw}, and we show the possible choices for $\alpha$ and 
$\theta=\alpha^2$ in Table~\ref{ZN}. Notice that, since we are interested
in models without untwisted tadpoles, we have already restricted
to even $N$ , $N=2P$.

As in the six-dimensional case, for even $N$ there exist two consistent
orientifold projections, which we denote of {\bf A}- and {\bf B}-type as
well, differing in their action on the order two twisted sector. The {\bf 
A}-type projection chooses left-right symmetric (antisymmetric) states in
the twisted NS-NS (RR) sector, while the {\bf B}-type projection does the
opposite. At the level of the massless closed string spectrum this amounts
to choosing either a $D=4$, $\NN=1$ chiral or a linear multiplet in the
corresponding sector. Given chiral-linear duality in four dimensions, both
projections lead to a similar closed string spectrum (even though in some
instances, like cancellation of $U(1)$ anomalies \cite{iru}, it is
important to distinguish chiral and linear multiplets \cite{abd, klein}). 
The closed string spectra for the models in Table~\ref{ZN} are provided in
Table~\ref{clospec}. Notice the unusual appearance of vector multiplets in
the untwisted sector. In geometrical terms, they arise from the reduction
of the type IIB self-dual 4-form, odd under $\Omega$, on $(2,1)$-forms odd
under the geometric action in $\Omega'\alpha$.

However, the two projections differ markedly in the twisted tadpoles they
generate, and therefore in the required open string spectra, as we discuss
in the following.

\begin{table}[ht]
\small
\renewcommand{\arraystretch}{1.25}
\begin{center}
\begin{tabular}{|c|c||c|c|}
\hline
$\IT^6/\IZ_N$ & $\Omega'\alpha$ twist & Untwisted & Twisted \\
\hline
$\IZ_2$   &${\tilde v}={1\over4}(1,1,-2)$ & 
6 {\bf V} + 8 {\bf Ch}  & 6 {\bf V} + 26 {\bf Ch} \\
\hline
$\IZ_4$   &${\tilde v}={1\over8}(1,3,-4)$ & 
4 {\bf V} + 6 {\bf Ch} & 8 {\bf V} + 28 {\bf Ch} \\
\hline
$\IZ_4$   &${\tilde v}={1\over8}(1,-3,2)$ & 
1 {\bf V} + 5 {\bf Ch} &  3 {\bf V} + 29 {\bf Ch} \\
\hline
$\IZ_{6}$ &${\tilde v}={1\over12}(1,-5,4)$ & 
5 {\bf Ch} & 2 {\bf V} + 27 {\bf Ch}  \\
\hline
$\IZ_{6}$ &${\tilde v}={1\over12}(1,5,-6)$ & 
4 {\bf V} + 6 {\bf Ch} &  8 {\bf V} + 28 {\bf Ch}\\
\hline
\end{tabular}
\caption{Closed string spectra for the $\Omega'\alpha$ orientifolds of
$\IT^6/\IZ_N$ orbifolds. We have not included the universally present
$\NN=1$ sugra multiplet and dilaton linear multiplet.}
\label{clospec}
\end{center}
\end{table}

\subsection{Models with {\bf A}-type projection}

The computation of the twisted crosscap tadpoles can be carried out by
following standard techniques. In the appendix, we compute these twisted
charges by merely adapting to our orientifold action the computations
for four-dimensional models in the Appendix of \cite{afiv}. We would like
to point out that, although our interest in the present paper is centered
on compact models, the calculation is valid for the general non-compact
case of $\Omega'\alpha$ orientifolds of $\IC^3/\IZ_N$, where one drops the
crystallographic condition on the twists. Hence, such models can be used
to define a large family of `twisted orientifold planes', uncharged with
respect to untwisted RR fields, but with non-zero twisted charges.
The crosscap tadpoles for this type of orientifold models are
\beqa
{\cal T}_{\theta^k} & = & - 16 \delta_{k,1\;{\rm mod}\; 2} [\,\sin(\pi k
v_1) + \sin(\pi k v_2)\,] \quad, \quad k=1,\ldots, N-1
\label{crosscap3}
\eeqa
Our twisted charges are normalized in units such that the 
$\theta^k$-twisted charges of D3- and D7$_i$-branes are $8\prod_{j=1}^3
\sin(\pi kv_j) \Tr \gamma_{\theta^k,3}$ and $2\sin(\pi k v_i)
\Tr\gamma_{\theta^k,7_i}$, respectively.

In the four-dimensional case, twisted tadpoles may have non-trivial
dependence with the compactification volumes, denoted $V_i$ for the
$i^{th}$ complex plane (we also introduce a regularized volume $V_4$ for
four-dimensional spacetime). Concretely, a crosscap twisted by
$\Omega'\alpha$ produces a tadpole proportional to $\sqrt{V_i}$ if
$\alpha^{2k+1}$ reflects the $i^{th}$ plane, and proportional to
$1/\sqrt{V_i}$ if it leaves $z_i$ invariant. A disk twisted by $\theta^k$
produces a tadpole proportional to $\sqrt{V_i}$ if $z_i$ is a Neumann
direction left invariant by $\theta^k$, and proportional to $1/\sqrt{V_i}$
if it $z_i$ is a Dirichlet direction left invariant by $\theta^k$. Any
tadpole is proportional to $\sqrt{V_4}$. These dependences are
implicitly taken into account in all tadpole cancellations in this paper.

In general, cancellation of twisted tadpoles will require the introduction
of open string sectors. The {\bf A}-type projection corresponds to a
choice of Chan-Paton matrices (\ref{cpone}) and (\ref{cptwosym}) for
D3-branes, (\ref{cpone}) and (\ref{cptwoanti}) for D7$_3$-branes, and
(\ref{cpthree}) and (\ref{cpfouranti}) for both D7$_1$- and D7$_2$-branes.
\footnote{This flip between the vector structure of different branes can
be shown as in \cite{pru}. Note the ambiguity in labeling the two
projections in the closed sector as with and without vector structure,
and the advantage of our labeling as {\bf A}- and {\bf B}-type.}.

There are several compact models which, however, do not lead to twisted
tadpoles. This is the case for the orientifolds with $\Omega'\alpha$
twists given by ${\tilde v}=\frac18(1,-3,2)$ and ${\tilde v}=\frac
1{12}(1,-5,4)$. These models do not require open string sectors, and
define consistent unoriented closed string theories. The corresponding
closed string spectra can be read from the relevant rows in 
Table~\ref{clospec}.

The remaining cases correspond to orientifolds with non-zero twisted
tadpoles, which we may try to cancel by the introduction of e.g. a suitable
set of branes and antibranes. We describe them in the following,

\medskip

{\noindent $\bullet$ {\bf The ${\tilde v}=\frac 14(1,1,-2)$ model}}

The contribution from the $\Omega'\alpha$, $\Omega'\alpha^3$ crosscaps to
the $\theta$-twisted tadpole (\ref{crosscap3}) is proportional to
$\sqrt{V_4V_3}$. This cannot be cancelled by using D3- or D7$_3$-branes,
but it can be cancelled by D7$_1$- or D7$_2$-branes. Here we describe a
particular solution to the tadpole conditions using an equal number of
D7$_1$- and \Dseven$_1$-branes. There are sixteen fixed points of
$\Omega'\alpha$, with coordinates $(z_1,z_2,z_3)$ with $z_1=0,(1+i)/2$,
$z_2=0, (1+i)/2$, $z_3=0,1/2,i/2,(1+i)/2$. At each of these points,
indexed by $P$, we need
\beqa
\Tr \gamma_{\theta,7_1,P} -\Tr \gamma_{\theta,{\bar 7}_1, P} \;  
- 16 = 0
\label{zcuatro}
\eeqa
There are also sixteen fixed tori of $\theta$, out of which four (with
coordinates $(z_1,z_2)$, with $z_1=0,(1+i)/2$, $z_2=0,(1+i)/2\,$) are fixed
under $\Omega'\alpha$, and twelve are not. At the latter, indexed by $Q$, 
$\theta$-twisted crosscap tadpoles are not generated, and hence the
corresponding twisted charges of the D7$_1$- and \Dseven$_1$-branes should
be zero.
\beqa
\Tr \gamma_{\theta,7_1,Q} - \Tr \gamma_{\theta,{\bar 7}_1, Q} = 0 
\eeqa

We consider the following solution. We locate 16 D7$_1$-branes at
$z_1=0$ in the first plane, and 16 \Dseven$_1$-branes at $z_1=(1+i)/2$,
with
\beqa
\gamma_{\theta,7_1,0}=\id_{16} \quad, \quad 
\gamma_{\theta,{\bar 7}_1,(1+i)/2}=-\id_{16} 
\eeqa
which satisfy the $\theta$-tadpole condition (\ref{zcuatro}) at the points
$P$. In order to avoid the contribution of twisted charges to the points 
$Q$ which they cross, we introduce Wilson lines along the second complex
plane
\beqa
\gamma_{W_2,7_1,0}=\diag(\id_8,-\id_8) \quad ,\quad
\gamma_{W_2,{\bar 7}_1,(1+i)/2}=\diag(\id_8,-\id_8) 
\eeqa
This completes the construction of the model. The resulting open
string spectrum is given in Table~\ref{fourone}. 

\begin{table}[ht]
\small
\renewcommand{\arraystretch}{1.25}
\begin{center}
\begin{tabular}{|c||c|}
\hline
${\bf 7}_1{\bf 7}_1$ & $\NN=1$ Vectors of $USp(8)^2$ \\
\hline
   & $\NN=1$ Chiral ({\bf 36},{\bf 1}) + ({\bf 1}, {\bf 36}) \\
\hline
${\bf{\bar 7}}_1{\bf{\bar 7}}_1$ & Gauge Bosons of $SO(8)^2$ \\
\hline
 & Fermion$_+ \;\;$ $({\bf 35}+{\bf 1},{\bf 1}) + ({\bf 1},{\bf 35}+
{\bf 1})$ \\
\hline
 & Cmplx. Scalars \ \ ({\bf 28},{\bf 1}) + ({\bf 1},{\bf 28}) \\
\hline
 & Fermion$_-\;\;$ $({\bf 35}+{\bf 1},{\bf 1}) + ({\bf 1},{\bf 35}+{\bf 
1})$ \\
\hline
\end{tabular}
\caption{Open string spectrum of the $\Omega'\alpha$ orientifold ${\tilde
v}=\frac 14(1,1,-2)$ of $\IT^4/\IZ_2$ with {\bf A}-type projection.}
\label{fourone}
\end{center}
\end{table}

\medskip

{\noindent $\bullet$ {\bf The ${\tilde v}=\frac 18(1,3,-4)$ model}}

In this model, the crosscaps associated to $\Omega'\alpha$ and 
$\Omega'\alpha^3$ produce a contribution to the $\theta$-,
$\theta^3$-twisted tadpoles proportional to $\sqrt{V_4V_3}$. As in the
previous model, they may be cancelled by the introduction of D7$_1$- or
D7$_2$-branes. In the following we construct a particular solution using
an equal number of D7$_1$- and \Dseven$_1$-branes. 

The model contains eight $\Omega'\alpha$-fixed points, at the locations
denoted by $(P_1,z_3)$, $(P_2,z_3)$, where $z_3=0,1/2,i/1,(1+i)/2$, and
$P_j$ denotes the location of the fixed points in the first two complex
planes, whose explicit form is not necessary here. At each of these eight
points, indexed by $P$, we need
\beqa
\Tr \gamma_{\theta,7_1,P} -\Tr \gamma_{\theta,{\bar 7}_1, P} \;  -16 = 0
\eeqa
There are also additional fixed points of $\theta$ which are not fixed
under $\Omega'\alpha$, and do not generate crosscap tadpoles. At these
points, indexed by $Q$, we need
\beqa
\Tr \gamma_{\theta,7_1,Q} -\Tr \gamma_{\theta,{\bar 7}_1, Q} = 0
\label{empty}
\eeqa
Finally, the closed string sector does not generate $\theta^2$-twisted
tadpoles, hence the disk contributions should also vanish. At all points
of $P$ or $Q$ type we need
\beqa
\Tr \gamma_{\theta^2,7_1}-\Tr \gamma_{\theta^2,{\bar 7}_1} & = & 0
\eeqa
We consider the following solution. We locate 32 D7$_1$-branes passing
through $P_1$ and 32 \Dseven$_1$-branes passing through $P_2$, with
\beqa
\gamma_{\theta,7_1,P_1} = \diag(\id_{16},e^{2\pi i\frac 14} \id_8, e^{2\pi
i\frac 34} \id_8) \quad ; \quad \gamma_{\theta, {\bar 7}_1,P_2} =
\diag(e^{2\pi i\frac 14} \id_8, e^{2\pi i\frac 24} \id_{16}, e^{2\pi
i\frac 34} \id_8) 
\eeqa
Notice that fixed points of $\theta$ not fixed under $\Omega'\alpha$ (i.e.
of type $Q$) remain empty and satisfy (\ref{empty}) automatically.
The resulting spectrum is shown in Table~\ref{fourtwo}. 

\begin{table}[ht]
\small
\renewcommand{\arraystretch}{1.25}
\begin{center}
\begin{tabular}{|c||c|}
\hline
${\bf 7}_1{\bf 7}_1$ & $\NN=1$ Vectors of $USp(16)\times U(8)$ \\
\hline
 & $\NN=1$ Chiral \ \ $({\bf 16},{\bf {\bar 8}}) + ({\bf 16},{\bf 8}) +
({\bf 136},{\bf 1}) + ({\bf 1}, {\bf {\rm Adj}})$ \\
\hline
${\bf{\bar 7}}_1{\bf{\bar 7}}_1$ & Gauge Bosons of $U(8)\times SO(16)$ \\
\hline
 & Fermion$_+ \;\;$ ({\bf Adj}, {\bf 1}) + ({\bf 1}, {\bf 135}+{\bf 1}) \\
\hline
 & Cmplx. Scalars \ \  $({\bf 8},{\bf 16}) + ({\bf \bar{8}},{\bf 16}) +
({\bf {\rm Adj}}, {\bf 1}) + ({\bf 1},{\bf 120})$ \\
\hline
 & Fermion$_- \;\;$ $({\bf 8},{\bf 16}) + ({\bf \bar{8}},{\bf 16}) +
({\bf {\rm Adj}},{\bf 1}) + ({\bf 1}, {\bf 135}+{\bf 1})$ \\
\hline
\end{tabular} 
\caption{Open string spectrum of the $\Omega'\alpha$ orientifold ${\tilde
v}=\frac 18(1,3,-4)$ of $\IT^4/\IZ_4$ with {\bf A}-type projection.}
\label{fourtwo}
\end{center}
\end{table}

\medskip

{\noindent $\bullet$ {\bf The ${\tilde v}=\frac{1}{12}(1,5,-6)$ model}}

In this model there are four $\Omega'\alpha$-fixed points, which we denote
by $(0,z_3)$, where $0$ represents the origin in the first two complex 
planes, and $z_3=0,1/2,i/1,(1+i)/2$ gives their location in the third.
At these points $P$ there are crosscap tadpoles for $\theta$- and
$\theta^3$-twisted fields, proportional to the volume factor
$\sqrt{V_4V_3}$. In order to cancel them we introduce an equal number of
D7$_1$- and \Dseven$_1$-branes, whose Chan-Paton matrices are then
constrained by
\beqa
\Tr \gamma_{\theta,7_1,P} -\Tr\gamma_{\theta,{\bar 7}_1,P}  -16 
& = & 0 \nonumber \\
\Tr \gamma_{\theta^3,7_1,P} -\Tr\gamma_{\theta^3,{\bar 7}_1,P}  -16
& = & 0 
\eeqa
In this model there are no additional $\theta$-fixed points. The only
remaining constraint is the cancellation of $\theta^2$-twisted disk
tadpoles (since there are no corresponding crosscap contribution). We have
\beqa
\Tr\gamma_{\theta^2,7_1,P} - \Tr\gamma_{\theta^2,{\bar 7}_1,P} & = & 0 
\eeqa
The simplest solution is to locate D7$_1$- and \Dseven$_1$-branes passing
through the origin, and with
\beqa
\gamma_{\theta,7_1} = \id_8 \quad ; \quad \gamma_{\theta,{\bar 7}_1} =
-\id_8 
\eeqa
Notice that since the Chan-Paton matrices above have no common eigenvalues, 
the tachyonic $7_1{\bar 7}_1$, ${\bar 7}_17_1$ groundstates are projected
out. The resulting spectrum is shown in Table~\ref{fourthree}
\begin{table}[ht]
\small
\renewcommand{\arraystretch}{1.25}
\begin{center}
\begin{tabular}{|c||c|}
\hline
${\bf 7}_1{\bf 7}_1$ & $\NN=1$ Vectors of $USp(8)$ \\
\hline
   & $\NN=1$ Chiral \ \ ${\bf 36}$ \\
\hline
${\bf{\bar 7}}_1{\bf{\bar 7}}_1$ & Gauge Bosons of $SO(8)$ \\
\hline
 & Fermion$_+ \;\;$ ${\bf 35}+{\bf 1}$ \\
\hline
 & Cmplx. Scalars \ \  ${\bf 28}$ \\
\hline
 & Fermion$_-$ \ \ ${\bf 35}+{\bf 1}$ \\
\hline
\end{tabular}
\caption{Spectrum of the $\Omega'\alpha$ orientifold ${\tilde v}=\frac
1{12}(1,5,-6)$ of $\IT^4/\IZ_6$ with {\bf A}-type projection.}
\label{fourthree}
\end{center}
\end{table}

\subsection{Models with {\bf B}-type projection}

The twisted crosscap tadpoles for a $\Omega'\alpha$ orientifold of
$\IC^3/\IZ_N$ with the {\bf B}-type projection are computed in the
appendix, leading to the result
\beqa
{\cal T}_{\theta^k} & = & + 16 \sin(\pi k v_3) \;
\delta_{k,1\;{\rm mod}\; 2}
\quad ,\quad k=1,\ldots, N-1
\label{crosscap4}
\eeqa
In some of the models in Table~\ref{ZN} all twisted tadpoles vanish.
Therefore they do not require open string sector, and define consistent
unoriented closed string theories in four dimensions. This is the
case for the models with $\Omega'\alpha$ twists ${\tilde v}=
\frac 14(1,1,-2)$, ${\tilde v}=\frac 18(1,3,-4)$, and ${\tilde v}=\frac
1{12}(1,5,-6)$. The corresponding closed string spectra can be read from
the relevant rows in Table~\ref{clospec}.

The remaining cases correspond to orientifolds with non-zero twisted
tadpoles, which may be cancelled by the introduction of e.g. a suitable
set of branes and antibranes. In the following we describe these models.

\medskip

{\noindent $\bullet$ {\bf The ${\tilde v}=\frac 18(1,-5,4)$ model}}

In this $\Omega'\alpha$ orientifold of $\IT^6/\IZ_8$, the only twisted
tadpoles in the compact model reside at the four $\alpha$ fixed points,
indexed by $P$. We consider a model in which they are cancelled by the
introduction of an equal number of D3- and \Dthree-branes. The conditions
of cancellation of the twisted RR tadpoles (\ref{crosscap4}) for $k=1$ 
read
\beqa
\Tr \gamma_{\theta,3,P} - \Tr \gamma_{\theta, {\bar 3},P} & = & 4
\eeqa
For $k=2$ the twisted crosscap tadpoles (\ref{crosscap4}) vanishes. Since
$\theta^2$ has a fixed plane, all D3- \Dthree-branes at points $P$ are
sources for the same $\theta^2$-twisted field. The condition
of cancellation of charge then reads
\beqa
\sum_P \, (\; \Tr \gamma_{\theta^2,3,P} \; - \; \Tr
\gamma_{\theta^2,{\bar 3},P}\; ) & = & 0
\eeqa
The simplest solution to these conditions is to locate, at two of the
$\Omega'\alpha$-fixed points, four D3-branes, and four \Dthree-branes at
each of the remaining two, with
\beqa
\gamma_{\theta,3} = \id_{4} \quad, \quad \gamma_{\theta,{\bar 3}}=-\id_4
\eeqa
The resulting spectrum is given in Table~\ref{four}

\begin{table}[ht]
\small
\renewcommand{\arraystretch}{1.25}
\begin{center}
\begin{tabular}{|c||c|}
\hline
${\bf 33}$ & $\NN=1$ Vectors of $SO(4)^2$ \\
\hline
${\bf{\bar 3}}{\bf{\bar 3}}$ & Gauge Bosons of $USp(4)^2$ \\
\hline
 & Fermion$_+$ \ \ $({\bf 5}+{\bf 1},{\bf 1}) + ({\bf 1},{\bf 5}+{\bf 1})$
\\
\hline
\end{tabular}
\caption{Open string spectrum of the $\Omega'\alpha$ orientifold (with
twist ${\tilde v}=\frac 18(1,-3,2)$) of $\IT^6/\IZ_4$ with {\bf B}-type
projection.}
\label{four}
\end{center}
\end{table}

\medskip

{\noindent $\bullet$ {\bf The ${\tilde v}=\frac 1{12}(1,-5,4)$ model}}

We conclude with the discussion of this $\Omega'\alpha$ orientifold of
$\IT^6/\IZ_6$, and by showing there is no configuration of branes and
antibranes leading cancellation of all twisted tadpoles. 

The argument goes as follows. First notice that we can treat branes and
antibranes in a unified manner by using only branes but formally allowing
the multiplicities of their Chan-Paton eigenvalues to take negative values.
The $\Omega'\alpha$ crosscaps generate a $\theta$-twisted tadpole at the
three $\alpha$-fixed points, which sit at the origin in the first two
planes, and at three points, labeled by $n=0,\pm 1$ in the third. We
introduce a general configuration of branes, passing through these points.
We include D7$_1$- and D7$_2$-branes passing through the origin in the
first two planes, and with arbitrary order three Wilson lines
$\gamma_{W_3,7_1}$, $\gamma_{W_3,7_2}$ in the third. We also consider
D7$_3$-branes sitting at the three different locations in the third plane,
and D3-branes, at the origin in the first two planes, and in the three
locations in the third. The $\theta$-twisted tadpole equation reads
\beqa
\frac{1}{\sqrt{3}} \Tr ( \gamma_{\theta, 7_1} \gamma_{W_3,7_1}^n ) -
\frac{1}{\sqrt{3}}\Tr ( \gamma_{\theta,7_2} \gamma_{W_3,7_2}^n ) +
\Tr \gamma_{\theta,7_3,n} - \Tr \gamma_{\theta,3,n} + 8 = 0
\label{abs1}
\eeqa

The crosscap twisted by $\Omega'\alpha^3$ generates a tadpole in the
$\theta^3$-twisted sector, proportional to the volume factor
$\sqrt{V_4/V_3}$. This $\theta^3$-tadpole also receives contributions from
disks twisted by $\theta^3$ associated with D3- and D7$_3$-brane, 
regardless of their location in the third plane. There is no contribution
with this volume dependence from D7$_1$- or D7$_2$-branes, hence the
$\theta^3$ tadpole condition reads
\beqa
\sum_n ( \Tr \gamma_{\theta^3,7_3,n} - 4\, \Tr \gamma_{\theta^3,3,n})+8=0 
\label{abs2}
\eeqa
We now show that (\ref{abs1}) and (\ref{abs2}) are incompatible. It is
enough to consider the equations mod 3. Let us sum (\ref{abs1}) over $n$, 
and take the resulting equation ${\rm mod} \; 3$. We obtain
\beqa
&  \frac{1}{\sqrt{3}} \Tr [\, \gamma_{\theta, 7_1} (\sum_{n=0,\pm 1}
\gamma_{W_3,7_1}^n )\, ] - \frac{1}{\sqrt{3}}\Tr [\,
\gamma_{\theta,7_2} (\sum_{n=0,\pm 1} \gamma_{W_3,7_2}^n )\, ] + &
\nonumber\\
& + \sum_{n=0,\pm 1} \Tr \gamma_{\theta,7_3,n} - \sum_{n=0,\pm 1}
\Tr \gamma_{\theta,3,n} = 0 \;\; {\rm mod}\; 3 & 
\label{mod}
\eeqa
Expressing the D3- Chan-Paton traces in terms of the eigenvalue
multiplicities from (\ref{cpthree}), we have
\beqa
\Tr \gamma_{\theta,3,n} = n_0 + n_1 - n_2 - n_3 = n_0 -2 n_1 + 2 n_2 - n_3
\; {\rm mod}\; 3 = \Tr \gamma_{\theta^3,3,n} \;{\rm mod}\; 3 \ 
\label{tmodthree}
\eeqa
and analogously for D7$_3$-branes. Also, since the Wilson line $W_3$ is
of order three, the eigenvalues in $\gamma_{W_3, 7_1}$ are $1$, $e^{\pm
2\pi i/3}$, and we can show $ \sum_n \gamma_{W_3,7_1}^n = 
0 \;\; {\rm mod}\; 3$. Finally, it is important to notice that the D7$_1$-
and D7$_2$-brane Chan-Paton matrices are of the form (\ref{cpone}),
without vector structure, hence their contributions $\tr\gamma_{\theta}$
in (\ref{mod}) are of the form an integer times $\sqrt{3}$, this latter
factor cancelling the coefficient $1/\sqrt{3}$.

Using these properties in (\ref{mod}), we get
\beqa
\sum_n ( \Tr \gamma_{\theta^3,7_3,n} - 4\, \Tr \gamma_{\theta^3,3,n}) = 0
\;\; {\rm mod} \; 3
\eeqa
in contradiction with (\ref{abs2}). Notice that there may be additional
contribution to (\ref{abs2}) from additional D3- or D7$_3$-branes sitting
at the origin in the first two planes, and at arbitrary points (not
$\theta$-fixed) in the third. However, these objects necessarily come in
three identical copies, permuted by the action of $\theta$, and their
contribution is zero ${\rm mod}\; 3$, and does not change the argument
above. We have not found any additional contributions which could help in
satisfying the tadpole conditions, and so have not succeeded in
constructing a consistent version of this orientifold model.

It is interesting to rephrase the problem we have encountered in a more
abstract language. Given an orientifold model, one can envision the space
of all possible (untwisted and twisted) charges under its RR fields, which
we loosely refer to as the charge lattice (even though some charges may
not be $\IZ$-valued). The charges of the orientifolds in the model define
a vector in that space, and the problem of cancelling the RR tadpoles
corresponds to constructing a state of the theory, typically
a configuration of branes, with precisely that charge vector. Clearly, a
problem arises if the charge vector defined by the crosscap charges
corresponds to an empty site in the charge lattice, i.e. a set of charges
for which no state in the theory exists.

This situation is precisely what we have found in the above model. We have
shown that for a very general kind of brane configurations the set of
states reproducing the correct $\theta$-twisted charges fills only a
subset of all possible $\theta^3$-twisted charges, namely zero ${\rm
mod}\; 3$. The crosscap contribution however, corresponds to a
$\theta^3$-twisted charge equal to $-2$, which is an empty site in the
sense explained above. Whether other kinds of branes can fill the empty
site, and provide a consistent completion of the model remains an open
question.

We conclude by pointing out that other orientifold models have been
previously claimed to be inconsistent due to uncancelable tadpoles.
In particular, the $\Omega$ orientifold of $\IT^6/\IZ_4$ with {\bf A}-type
projection \cite{afiv} was shown to contain a twisted tadpole with an
unusual volume dependence, and whose cancellation could not be achieved
with the D-branes in the model. In Section 5 we review this problem, and
find that in fact it is possible to construct a set of brane-antibrane
pairs canceling precisely this problematic tadpole, and rendering the
model consistent. This nicely illustrates the power of non-BPS
configurations of branes in cancelling twisted charges. Our above example
in this Section, though, suggests there still exist orientifolds for
which tadpole cancellation cannot be achieved.

\section{Further six-dimensional models}

In this section we study a set of six-dimensional models without untwisted
tadpoles, obtained by modding out type IIB theory on $\IT^4/\IZ_N$, for
even $N$, by the orientifold action $\Omega\Pi$, with $\Pi:(z_1,z_2)\to
(z_2,-z_1)$. Notice that closure requires $N$ to be even, $N=2P$, therefore 
it is possible to construct compact $\Omega\Pi$ orientifolds of the
$\IZ_2$, $\IZ_4$ and $\IZ_6$ orbifolds.

This type of orientifold projection has been studied in \cite{upermu}
in the non-compact context, i.e. orientifolds of $\IC^2/\IZ_N$, where the 
crystallographic constraint can be dropped.The closed string twisted
sector, for a $\IC^2/\IZ_N$ point gives rise to $N-1$ tensor multiplets of
$D=6$ $\NN=1$ supersymmetry. The closed string sector contribution to
twisted tadpoles can be computed to be
\beqa
{\cal T}_{\theta^k}& = & - 16 N \delta_{k,N/2}\; = 0
\eeqa
in units where the twisted charges of D5- and D9-branes are $4\sin^2(\pi 
k/N) \Tr\gamma_{\theta^k,5}$ and $\Tr\gamma_{\theta^k,9}$ respectively.

Since there is always at least one non-zero twisted tadpole, the
configurations requires open string sectors for consistency. We will
introduce sets of D5-branes and \Dfive-branes in order to cancel twisted
tadpoles (D9- and \Dnine-branes may also be easily incorporated). The
general structure of their Chan-Paton matrices is 
\beqa
\gamma_{\theta^k,5} & = & \diag(\id_{n_0}, e^{2\pi i\frac 1N} \id_{n_1},
\ldots, e^{2\pi i\frac {N-1}{N}} \id_{n_{N-1}} ) \nonumber \\
\gamma_{\Omega\Pi,5} & = & \diag(\id_{n_0}, \varepsilon_{n_1}, \ldots,
\id_{n_{N-2}}, \varepsilon_{n_{N-1}})
\eeqa
and analogously for \Dfive-branes. The open string sector, for instance in
the absence of \Dfive-branes, gives rise to the following multiplets on
the D5-brane world-volume
\beqa
{\rm Vector} & SO(n_0)\times USp(n_1)\times\ldots\times SO(n_{N-2})\times
USp(n_{N-1}) \nonumber \\
{\rm Hyper} & \frac 12 (n_0,n_1) + \frac 12 (n_1,n_2) + \ldots +
\frac 12 (n_{N-1},n_0)
\eeqa
If instead we have only \Dfive-branes and no D5-branes, we obtain a
non-supersymmetric version of the above spectrum, with the only
difference that the `gauginos' transform in the `wrong' two-index
representation of the gauge group (symmetric of $SO(n_{2i})$ and
antisymmetric of $USp(n_{2i+1})$).

\medskip

Let us turn to the construction of compact models. The $\Omega\Pi$
orientifold of the $\IT^4/\IZ_2$ orbifold is actually equivalent to the
$\Omega\alpha$ orientifold of the $\IT^4/\IZ_2$ orbifold studied in
Section 2, as can be shown by diagonalizing $\Pi$ without changing the
structure of the orbifold twist $\theta$. Hence it does not give rise to a
new model \footnote{Interestingly, in the non-compact contex the
$\Omega\alpha$ and $\Omega\Pi$ orientifold of $\IC^2/\IZ_2$ are T-dual to
two different configurations (in the spirit of \cite{hw}) of type IIA
NS-fivebranes, D-branes and orientifold planes, studied in \cite{lll} and
\cite{towards}, which nevertheless give the same massless spectrum.}. 
The situation is different for the $\IZ_4$ and $\IZ_6$ orbifolds, since
diagonalizing $\Pi$ changes the action of $\theta$.

Let us center on the $\Omega\Pi$ orientifold of $\IT^4/\IZ_4$. In the
untwisted closed string sector, we obtain the $D=6$ $\NN=1$ gravity 
multiplet, the dilaton tensor multiplet, and two hyper- and two tensor
multiplets. In the twisted sector, two of the four $\theta$-fixed points
are also fixed under $\Omega\Pi$, and produce three tensor multiplets
each. The remaining two points are exchanged by $\Omega\Pi$, and give a
total of three hyper- and three tensor multiplets. There are also twelve
fixed points of $\theta^2$ which are grouped in pairs by the action of
$\theta$, i.e. they correspond to six $\IC^2/\IZ_2$ singularities in the
quotient. Two of them are invariant under $\Omega\Pi$, and give one tensor
multiplet each, while the remaining four produce a total of two hyper- and
two tensor multiplets. In total, we have six hyper- and fourteen tensor
multiplets.

Points fixed under $\Omega\Pi$ generate twisted tadpoles. In order to
cancel them, we introduce e.g. 16 D5-branes at one of the points fixed
under $\theta$ and $\Omega\Pi$ and 16 \Dfive-branes at the other, with
\beqa
\gamma_{\theta,5,P_1}=\diag(\id_8,e^{2\pi i \frac 24}\id_8) \quad ; \quad
\gamma_{\theta,{\ov 5},P_2}=\diag(e^{2\pi i \frac 14}\id_8,e^{2\pi
i\frac 34}\id_8) 
\eeqa
Also, we achieve cancellation of twisted tadpoles at the two $\IC^2/\IZ_2$
points fixed under $\Omega\Pi$, by locating eight D5-branes at one of
them, and eight \Dfive-branes at the other, with $\gamma_{\theta^2,5,Q_1}=
\id_8$, $\gamma_{\theta^2,{\ov 5},Q_2}=-\id_8$. The complete spectrum for
this configuration, shown in Table~\ref{sixthree}, is free of gauge and
gravitational anomalies.

\begin{table}[ht]
\small
\renewcommand{\arraystretch}{1.25}
\begin{center}
\begin{tabular}{|c|c||c|}
\hline
{\bf Closed} & Untwisted & $\NN=1$ SUGRA + 3 {\bf T} + 2{\bf H} \\
\hline
             & Twisted & 12 {\bf T} + 4 {\bf H} \\
\hline\hline
{\bf Open} & {\bf 55} & $\NN=1$ Vectors of $[SO(8)^2]_{\IZ_4}\times
SO(8)_{\IZ_2}$ \\ 
\hline
           &  ${\bf{\bar 5}}{\bf{\bar 5}}$ & Gauge bosons of
$[USp(8)^2]_{\IZ_4}\times USp(8)_{\IZ_2}$ \\ 
\hline
           & & Fermion$_+$ $({\bf 27}+{\bf 1}, {\bf 1}; {\bf 1}) + ({\bf 
1},{\bf 27}+{\bf 1}; {\bf 1}) + ({\bf 1}, {\bf 1}; {\bf 27}+{\bf 1})$ \\ 
\hline
\end{tabular} 
\caption{Spectrum of the $\Omega\Pi$ orientifold of $\IT^4/\IZ_4$.}
\label{sixthree}
\end{center}
\end{table}

\section{Tadpole cancellation in the $\IT^6/\IZ_4$ orientifold}

In this section we use the philosophy developed in our understanding of
twisted tadpoles to solve and old puzzle in the construction and tadpole
cancellation in the $\Omega$ orientifold of $\IT^6/\IZ_4$, with the {\bf
A}-type projection (usually called without vector structure) \cite{afiv}. 
Here the $\IZ_4$ is generated by an action $\theta$ with twist $v=\frac
14(1,1,-2)$.  Even though this model, which contains untwisted tadpoles,
appears unrelated to our main interest in the present paper, there is an
interesting relation, based on the structure of twisted tadpoles in the
$\IZ_4$ orientifold. Specifically, it contains a twisted tadpole whose
cancellation is possible only by a set of D-branes (and antibranes) 
completely unrelated to the untwisted tadpoles of the model. The fact that
the twisted tadpole is cancelable in this manner had been overlooked
(mainly due to the fact that the introduction of antibranes in orientifold
constructions is relatively recent), leading to the (incorrect) claim that
the $\IZ_4$ orientifold, with the {\bf A}-type projection is inconsistent.

Let us start reviewing the tadpoles generated by the closed string sector.
and their volume dependence \cite{afiv}. For convenience, we prefer to
discuss the model in a T-dual version where the orientifold action is
$\Omega_3=\Omega R_3 (-)^{F_L}$. There are tadpoles for untwisted 
fields, generated by $\Omega_3$ and by $\Omega_3\theta^2$, which are
proportional to the volumes $(V_4V_1V_2/V_3)^{1/2}$ and $[V_4/(V_1 V_2
V_3)]^{1/2}$, respectively. They are cancelled by the introduction 32
D7$_3$- and 32 D3-branes, whose untwisted disk diagrams lead to charges
with the required volume dependence and coefficient.

The $\Omega_3\theta$ twisted crosscap leads to a tadpole proportional to
$\sqrt{V_4 V_3}$. As pointed out in \cite{afiv,kr2}, this twisted tadpole
cannot cancelled by any disk associated to the D7$_3$- or the D3-branes in
the model. However, by now we are familiar with the fact that some
orientifolds require specific sets of branes, designed to cancel twisted
tadpoles, even if untwisted tadpoles do not require them. With this
philosophy in mind, we introduce a set of D7$_1$-\Dseven$_1$ pairs
(one may introduce D7$_2$-\Dseven$_2$ pairs if desired), at the points
$P$ fixed under $\theta$, which have coordinates $(z_1,z_2,z_3)$, with 
$z_1=0,(1+i)/2$, $z_2=0,(1+i)/2$, $z_3=0,1/2,i/2, (1+i)/2$. The
$\theta^2$-twisted tadpole condition reads
\beqa
\Tr \gamma_{\theta^2,7_1,P} - \Tr \gamma_{\theta^2,{\bar 7}_1,P} & = & -16
\eeqa
At points $Q$ fixed under $\theta^2$ but not under $\theta$, the condition
must be
\beqa
\Tr \gamma_{\theta^2,7_1,Q} - \Tr \gamma_{\theta^2,{\bar 7}_1,Q} & = & 0
\eeqa
The closed string sector does not generated tadpoles for $\theta$-twisted
fields. The cancellation of twisted disk contributions requires
\beqa
\sqrt{2} \, (\; \Tr \gamma_{\theta, 7_1,P} - \Tr\gamma_{\theta,{\bar 7}_1,
P} \; )\, - 2 \Tr \gamma_{\theta, 7_3, P} -4 \Tr \gamma_{\theta,3} & = & 0
\label{tad}
\eeqa

At this point, we would like to recall that for the {\bf A}-type
projection, D7$_3$- and D3-branes have Chan-Paton matrices with vector
structure (roughly speaking $\gamma_{\theta}^N=-\id$) and D7$_1$-,
\Dseven$_1$-branes (and also D7$_2$-, \Dseven$_2$-branes if introduced)
have matrices with vector structure ($\gamma_{\theta}^N=+\id$).

It is relatively easy to build models satisfying all tadpole conditions.
To give an example, we proceed as follows. In order to cancel the
problematic $\theta^2$-twisted tadpoles, we introduce 16 D7$_1$ branes at
$z_1=0$, and 16 \Dseven$_1$-branes at $z_1=(1+i)2$, with 
\beqa
\gamma_{\theta,7_1,0}=\diag(e^{2\pi i\frac 14}\id_8, e^{2\pi i\frac 
34}\id_8) \quad , \quad  \gamma_{\theta,{\bar 7}_1,(1+i)/2}=\diag(\id_{8},
-\id_8)
\eeqa
In order to avoid generating $\theta^2$-twisted tadpoles at points $Q$,
not fixed under $\theta$, we introduce Wilson lines along the second
complex plane, acting on the D7$_1$-, \Dseven$_1$-branes as
\beqa
\gamma_{W_2,7_1,0} = \diag(\id_4,-\id_4,\id_4,-\id_4) \quad , \quad
\gamma_{W_2,{\bar 7}_1,(1+i)/2} = \diag(\id_4,-\id_4,\id_4,-\id_4)
\eeqa
Finally, we need to introduce the set of 32 D7$_3$- and 32 D3-branes to
cancel the untwisted tadpoles, and constrained by (\ref{tad}). A simple
choice is just
\beqa
\gamma_{\theta,7_3} = \gamma_{\theta,3} = (e^{\pi i\frac 14} \id_8, e^{\pi
i\frac 34} \id_8, e^{\pi i\frac 54}\id_8, e^{\pi i \frac 78}\id_8)
\eeqa
with all D7$_3$-branes sitting for instance at $z_3=0$, and all D3-branes
sitting at $(0,0,0)$. For completeness we also provide the Chan-Paton
matrices for $\Omega_3$
\beqa
\begin{array}{lcl}
\gamma_{\Omega_3,7_3}={\pmatrix{
& & & \id_8 \cr
& & \id_8 & \cr
& \id_8 & & \cr
\id_8 & & & \cr
}} & \quad  ; \quad & 
\gamma_{\Omega_3,3} = {\pmatrix{
& & & \id_8 \cr
& & \id_8 & \cr
& -\id_8 & & \cr
-\id_8 & & & \cr
}} \\
\gamma_{\Omega_3,7_1}={\pmatrix{
& \id_8 \cr
-\id_8 & \cr}} 
& \quad  ; \quad &\gamma_{\Omega_3,{\bar 7}_1}=
{\pmatrix{
\varepsilon_8 & \cr
& \varepsilon_8 \cr}}
\end{array}
\eeqa
With this information it is a simple exercise to compute the resulting
spectrum, which is given in Table~\ref{tachan}. It is interesting to point
out that for instance the cancellation of non-abelian anomalies for gauge
factors on the D3-branes is crucially correlated with the cancellation of
the problematic twisted tadpole. 

Clearly, the lesson we have learned about the consistency of the
$\IT^6/\IZ_4$ orientifold extends to other cases previously considered
inconsistent as well, like orientifolds (with {\bf B}-type projection) of
$\IT^6/\IZ_8$, $\IT^6/\IZ_8'$, and $\IT^6/\IZ_{12}'$, which contain the
same type of twisted tadpole \cite{afiv}. Hoping that the $\IZ_4$ example
above suffices to illustrate the idea, we leave further issues for future
work.
\begin{table}[ht]
\footnotesize
\renewcommand{\arraystretch}{1.25}
\begin{center}
\begin{tabular}{|c|c||c|}
\hline
{\bf Closed} & Untwisted & $\NN=1$ SUGRA + 1 {\bf L} + 6 {\bf C}  \\
\hline
             & Twisted &  32 {\bf {Ch}} \\
\hline\hline
{\bf Open} & ${\bf 33}$ & $\NN=1$ Vectors of $U(8)^2$ \\
\hline
                 &          & $\NN=1$ Chiral \ $2\; \times\;
[\,({\bf 8},{\bf {\bar 8}}) + ({\bf 1},{\bf 28}) + ({\bf
{\ov{28}}},{\bf 1})\, ] + ({\bf 8},{\bf 8}) + ({\bf\bar{8}},{\bf\bar{8}})$
\\
\hline
                 & ${\bf {7_3 7_3}}$ & $\NN=1$ Vectors of $U(8)^2$ \\
\hline
                 &          & $\NN=1$ Chiral\ $2\; \times\;
[\,({\bf 8},{\bf{\bar 8}}) + ({\bf 1},{\bf 28}) + ({\bf{\ov{28}}},{\bf 
1})\, ] + ({\bf 8},{\bf 8}) + ({\bf\bar{8}},{\bf\bar{8}})$
\\
\hline
                 & ${\bf {7_1 7_1}}$ & $\NN=1$ Vectors of $U(4)^2$ \\
\hline
                 & & $\NN=1$ Chiral\ $2\ \times \ [\,({\bf 6},{\bf 1}) + 
({\bf{\ov 6}},{\bf 1}) + ({\bf 1},{\bf 6}) + ({\bf 1},{\bf {\ov 6}})\, ] $
\\
\hline
           & ${\bf{\bar 7}_1}{\bf{\bar 7}_1}$ & Gauge Bosons of
$USp(4)^4$ \\
\hline
           & & Fermion$_+$ \ \ $({\bf 5}+{\bf 1},{\bf 1},{\bf 1},{\bf 1})
+ ({\bf 1},{\bf 5}+{\bf 1},{\bf 1},{\bf 1}) + ({\bf 1},{\bf 1},{\bf
5}+{\bf 1},{\bf 1}) + ({\bf 1},{\bf 1},{\bf 1},{\bf 5}+{\bf 1}) $ \\
\hline
           & & Cmplx. Scalars \ \ $({\bf 4},{\bf 1},{\bf 4},{\bf 1}) +
({\bf 1},{\bf 4},{\bf 1},{\bf 4})$ \\
\hline
           & & Fermion$_-$ \ \ $({\bf 4},{\bf 1},{\bf 4},{\bf 1}) +
({\bf 1},{\bf 4},{\bf 1},{\bf 4})$ \\
\hline
           & ${\bf{37_3}}+ {\bf 7_3 3}$ & $\NN=1$ Chiral
$({\bf 8},{\bf 1};{\bf 1},{\bf{\bar 8}}) + ({\bf 1},{\bf 8};{\bf 1},{\bf 
8}) + ({\bf 1},{\bf{\bar 8}};{\bf 8},{\bf 1}) + ({\bf{\bar 8}},{\bf 
1};{\bf {\bar 8}},{\bf 1})$ \\
\hline
           & ${\bf{37_1}}+ {\bf 7_1 3}$ & $\NN=1$ Chiral
$({\bf {\bar 8}},{\bf 1};{\bf 4},{\bf 1})+ ({\bf {\bar 8}},{\bf
1};{\bf 1},{\bf 4}) + ({\bf 1},{\bf 8};{\bf 4},{\bf 1}) + ({\bf
1},{\bf 8},{\bf 1},{\bf 4})$ \\ \hline
           & ${\bf{7_3 7_1}}+ {\bf 7_1 7_3}$ & $\NN=1$ Chiral
$({\bf {\bar 8}},{\bf 1};{\bf 4},{\bf 1}) + ({\bf 1},{\bf 8};{\bf
4},{\bf 1})$ \\
\hline
           & ${\bf{7_3 {\bar 7}_1}}+ {\bf {\bar 7}_1 7_3}$ & 
Fermion$_+$ \ \ $({\bf 8},{\bf 1};{\bf 4},{\bf 1},{\bf 1},{\bf 1}) +
({\bf 1},{\bf {\bar 8}};{\bf 1},{\bf 1},{\bf 4},{\bf 1})$ \\
\hline
& & Cmplx. Scalars \ \ $({\bf 8},{\bf 1};{\bf 4},{\bf 1},{\bf 1},{\bf 1})
+ ({\bf 1},{\bf{\bar 8}};{\bf 1},{\bf 1},{\bf 4},{\bf 1})$ \\
\hline
\end{tabular} 
\caption{Spectrum of a consistent $\Omega_3$ orientifold of $\IT^6/\IZ_4$
with {\bf A}-type projection.}
\label{tachan}
\end{center}
\end{table}

\section{Phase transitions and stability regions}

The non-supersymmetric models we have been considering are absolutely
stable against decay to a supersymmetric vacuum, because there is no
consistent (tadpole-free) supersymmetric vacuum for these orientifold
projections. However, in particular regions of parameter space the
configuration of branes and antibranes may become unstable and decay to
other (necessarily non-BPS) systems of branes. The analysis of such
decays and the stability regimes is analogous to that performed in the
study of stable non-BPS states in string theory. In this section we
address this issue in the simplest model we have constructed, namely the
$\Omega'\alpha$ orientifold of $\IT^4/\IZ_2$ with {\bf B}-type projection,
studied in Section~2. We will take advantage of by now standard properties
of brane-antibrane systems, and non-BPS branes, in $\IZ_2$ orbifolds (see
e.g. \cite{senkink, senvortex, stefanski} and references therein).

\begin{figure}
\begin{center}
\centering
\epsfysize=4cm
\leavevmode 
\epsfbox{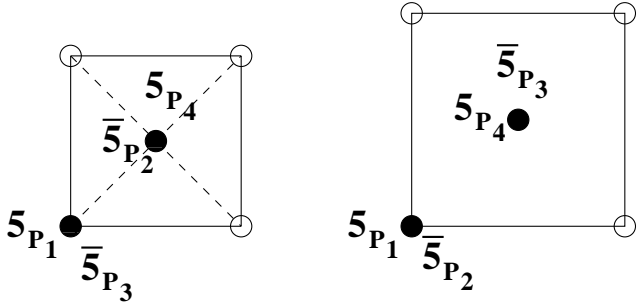}
\end{center}
\caption[]{\small Configuration of D5- and \Dfive-branes in the
$\Omega'\alpha$ orientifold of $\IT^4/\IZ_2$, for $R_1=1$, $R_2>1$. 
Discontinuous lines denote strings stretched between D5-branes at $P_1$
and \Dfive-branes at $P_2$, which lead to tachyonic modes for $R_1<1$.}
\label{five}  
\end{figure}

Let us consider the $\Omega'\alpha$ orientifold of $\IT^4/\IZ_2$. Denoting
the radii of the corresponding two-tori $R_1$, $R_2$, there are four fixed
points under the $\Omega'\alpha$ action, given by $(z_1,z_2)$ equal to
$P_1=(0,0)$, $P_2=(\frac{1+i}{2} R_1, 0)$, $P_3=(0,\frac{1+i}{2} R_2)$ and
$P_4=(\frac{1+i}{2} R_1, \frac{1+i}{2} R_2)$. Let us consider locating
8 D5-branes at $P_1$ and $P_4$ and 8 \Dfive-branes at $P_2$ and $P_3$,
with $\gamma_{\theta,5}=\id_8$ and $\gamma_{\theta, {\ov 5}}=-\id_8$. 
Table~\ref{sixtwo} provides the spectrum for this configuration, which is
depicted in Figure~\ref{five}. The configurations is symmetric under the
exchange of the first and second complex planes, a fact which simplifies
the analysis of the model.

The configuration is stable when $R_1$ and $R_2$ are large. But it may
become unstable against decay to some other non-supersymmetric
configuration if these geometric data change. In particular, the mass of a
string stretching between the D5-branes at $P_1$ and the \Dfive-branes at
$P_2$ (or between D5-branes at $P_4$ and \Dfive-branes at $P_3$) is 
\beqa
M^2 & = & \frac{1}{4\pi^2} (\pi\sqrt{2} R_1)^2 - 1/2
\eeqa
in $\alpha'=1$ units.
At $R_1=1$ the $5{\bar 5}$, ${\bar 5}5$ groundstates become massless, and
for $R_1<1$ they are tachyonic, and the configuration must decay to some
other non-supersymmetric configuration, with the same charges (i.e. same
contribution to the twisted and untwisted tadpoles). Let us center on this
phase transition, keeping $R_2$ large throughout the analysis. 

In the absence of the orientifold projection, the situation is reminiscent
of that studied in \cite{senkink}, namely a set of branes and antibranes,
sitting at the two fixed points in a circle modded out by a $\IZ_2$
action. When the fixed points come to a critical distance of $\sqrt{2}$
the tachyonic groundstate of stretched strings triggers a decay to a
non-BPS D-brane wrapped on the circle. In our context, the critical
distance between the fixed points is reached for $R_1=1$, so we may expect
a decay to a set of non-BPS D6-branes (denoted \Dtsix-branes) wrapped on
cycles passing through $z_1=0$ and $z_1=(1+i)/2$. 

In fact, since the branes and antibranes actually sit in a two-torus
rather than in a circle, there are more tachyons than those in
\cite{senkink}. We obtain tachyonic modes from strings stretching between
branes and antibranes (and viceversa) in the two different directions
depicted in Figure~\ref{five}. Therefore we can expect the decay to a set
of \Dtsix-branes will not be enough to stabilize the configuration for
$R_1<1$. In this respect, it is useful to describe the \Dtsix-brane
configuration quite explicitly, and discuss its set of light fields. We
will find it indeed contains tachyonic modes for $R_1<1$.

\begin{figure}
\begin{center}
\centering
\epsfysize=4cm
\leavevmode 
\epsfbox{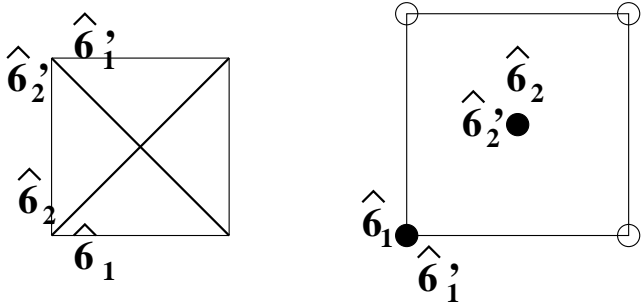}
\end{center}
\caption[]{\small Configuration of \Dtsix-branes in the $\Omega'\alpha$
orientifold of $\IT^4/\IZ_2$, for $R_1=1$, $R_2>1$.}
\label{six}  
\end{figure}

Hence we consider a set of \Dtsix-branes, denoted \Dtsix$_2$-branes, 
sitting at $z_2=0$, and partially wrapped on the first plane, so that they
pass through the points $(0,0)$, $(\frac{1+i}2 R_1,0)$, and another set of
\Dtsix-branes, denoted \Dtsix$_2$-branes, sitting at
$z_2=\frac{1+i}2 R_2$, and partially wrapped on the first plane so that
they pass through $(0,\frac{1+i}2 R_2)$, $(\frac{1+i}2 R_1,\frac{1+i}2 R_2)$.
We also need to introduce their $\Omega'\alpha$ images, denoted with a
prime. The complete configuration is depicted in Figure~\ref{six}.

In order for this configuration to have the same twisted charges as the
previous one, we must take $\gamma_{\hat 6_1}=\id_4$, $\gamma_{\hat 6_2}=
\id_4$. The contribution of \Dtsix-branes and \Dtsix$'$-branes to the
twisted tadpoles at each fixed point add up and cancel the contribution
from the closed string sector. 

Notice that the relation between tensions of the fractional D5-,
\Dfive-branes, and the truncated \Dtsix-branes is \cite{stefanski,
senkink} $T_{\hat 6}= \frac{\sqrt{2}}{2\pi} T_{5}$. The total tension of
the 8 six-branes after compactification on the $\IT^4$ is $8\times
2\pi\times \sqrt{2} R_1 \times T_{\hat 6} \ = \ 16 R_1 T_{5}$, and for
$R_1=1$ agrees with the tension of the 8 D5- plus 8 \Dfive-branes in the
initial configuration (Fig.~\ref{five}), making the transition
energetically possible. 

However, the configuration of \Dtsix-branes is actually unstable for 
$R_1<1$ due to the tachyonic groundstate of e.g. a $\hat{6}_1\hat{6}_1$
string stretching between the  points $(\frac {1+i}2 R_1,0)$ and $(R_1,0)$. 
Also, it is unstable for $R_1>1$ due to a tachyonic momentum mode e.g.
along the \Dtsix$_1$-brane. In any event, it is an interesting exercise to
obtain the chiral spectrum on this set of \Dtsix-branes, which is
marginally tachyon-free for $R_1=1$, and verify the cancellation of
anomalies in the effective six-dimensional field theory.

The spectrum in the $\hat{6}_1\hat{6}_1$ sector can be extracted from the
computations in \cite{senkink}. Notice that $\Omega'\alpha$ maps this
sector to the $\hat{6}_1'\hat{6}_1'$, so we may keep just the former and
not impose the orientifold projection. We obtain gauge bosons of $U(4)$,
and two right-handed fermions in the adjoint representation. We also need
to compute the $\hat{6}_1\hat{6}_1'$ sector, which is mapped to itself
under $\Omega'\alpha$, and so suffers the orientifold projection. An
important observation is that, since these branes intersect twice, the
spectrum appears in two copies. In the NS sector, each intersection would
give rise to tachyonic scalars, but they are projected out by the $\IZ_2$
orbifold projection \footnote{These tachyons would parametrize the
possibility of recombining the two intersecting \Dtsix-branes into smooth
ones, which would no longer pass through the fixed point. This is clearly
forbidden by conservation of $\IZ_2$ twisted charges, so the $\IZ_2$
projection removes the tachyons.}. In the R sector, the orbifold projection 
leaves one right-handed fermion (per intersection) transforming in the
adjoint of $U(4)$, and the orientifold projection reduces it to the 
two-index antisymmetric representation ${\bf 6}$. The $\hat{6}_1'
\hat{6}_1$ sector also gives two right-handed fermions in the ${\bf 6}$.
Finally, \Dtsix$_2$- and \Dtsix$_2'$ branes lead to the analogous spectrum. 
The complete open string spectrum is provided in Table~\ref{tablesix}. The
model is free of gauge and gravitational anomalies. Recall that, as
mentioned above, stretched strings and momentum modes lead to an
additional set of marginally massless scalars, not shown in
Table~\ref{tablesix}, which become tachyonic/massive at $R_1\neq 1$.
\begin{table}[ht]
\small
\renewcommand{\arraystretch}{1.25}
\begin{center}
\begin{tabular}{|c||c|}
\hline
${\bf{66}},\ {\bf{6' 6'}}$ & Gauge bosons of $U(4)^2$ \\ 
\hline
                & Fermion$_+ \ 2\times\ [ ({\rm Adj},1) + (1,{\rm Adj})] $
\\ 
\hline
 ${\bf{6 6'}}$ & Fermion$_+ \ 2\times\ [ (6,1) + (1,6) ]$ \\
\hline
 ${\bf{6' 6}}$ & Fermion$_+\  2\times\ [ (6,1) + (1,6) ]$ \\
\hline
\end{tabular}
\caption{Open string spectrum for the configuration of \Dtsix-branes in
Figure~\ref{six}.}
\label{tablesix}
\end{center}
\end{table}

The tachyonic momentum modes present at $R_1>1$ clearly signal an 
instability against decay to the original configuration of D5-\Dfive pairs. 
Let us instead consider the regime $R_1\leq1$, and look for a 
non-BPS configuration stable in that situation. Before the orientifold
projection, the initial configuration consists of a set of branes and
antibranes at fixed points in a square $\IT^2/\IZ_2$. The phase transition
when the size of the two-torus shrinks, is analogous to that studied in
\cite{senvortex}. In our context, we would obtain a set of D7-\Dseven
pairs wrapped on the first plane, and with suitable order two Wilson lines
turned on in order to reproduce the correct charges at the fixed points.
At this point a question arises, since the orientifold projection
$\Omega'\alpha$ projects out the 8-form under which D7-branes are charged,
so it seems that these should not be allowed in the model. The resolution
is simply that the orientifold projection actually maps D7-branes to
\Dseven-branes and viceversa, so that the object formed by a pair of
D7-\Dseven branes exchanged by the orientifold projection, is indeed an
allowed state in the orientifolded theory. Namely this state is neutral
under the 8-form, so its existence is consistent with the latter being
projected out. This non-BPS \Dtseven-brane is analogous to the non-BPS
D7-brane of type I string theory \cite{wittenk} (but not suffering
its stability problems \cite{lerda}, due to the absence of background
branes in the model).

\begin{figure}
\begin{center}
\centering
\epsfysize=4cm
\leavevmode 
\epsfbox{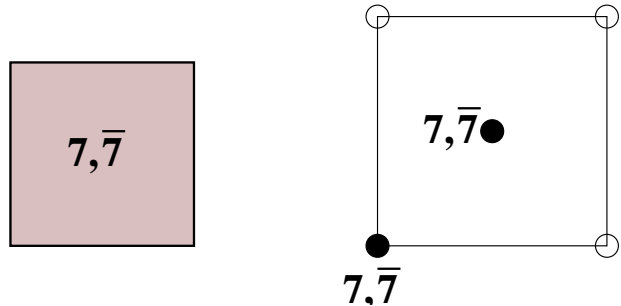}
\end{center}
\caption[]{\small Configuration of D7-\Dseven pairs in the $\Omega'\alpha$
orientifold of $\IT^4/\IZ_2$, for $R_1=1$. The branes wrap the first
(shaded) complex plane, and sit at the dot locations in the second.}
\label{seven}  
\end{figure}

Hence the initial configuration of D5-\Dfive branes decays, for $R_1<1$, to
two sets of D7-\Dseven pairs (or \Dtseven-branes), wrapped on the first
plane, and sitting at the locations $z_2=0$ and $z_2=\frac{(1+i)}{2} R_2$
in the second. The configuration is schematically shown in figure
\ref{seven}.

In order to reproduce the correct twisted charges at the $\alpha$-fixed
points, we need to take
\beqa
\gamma_{\theta,7_1}= e^{\pi i\frac 12}\id_8 \quad ; \quad \gamma_{\theta,
{\bar 7}_1} = e^{\pi i\frac 32}\id_8
\eeqa
Notice that D7-branes have Chan-Paton matrices `without vector structure'
\footnote{This can be derived by analyzing the $\IZ_2$ projection in 57
sectors, which exist at $R_1=1$ if for instance only part of the D5-\Dfive
branes are transformed into \Dtseven-branes. The result then follows for
$R_1<1$.}. In order to have zero contribution to the tadpoles at
$\theta$-fixed points not fixed under $\Omega'\alpha$, we introduce the
Wilson lines
\beqa
\gamma_{W_1,7} = \diag(\id_4,-\id_4) \quad ; \quad \gamma_{W_1,{\bar 7}} =
\diag(\id_4,-\id_4)
\eeqa

The tension of D7-branes is related to that of D5-branes by $T_{7}=
\frac{1}{4\pi^2} T_{5}$ \cite{dbrane}, so the total tension of the
sevenbranes, after compactification on the torus is $16 \times
(2\pi R_1)^2\ T_7 \ =\ 16 \times R_1^2\ T_{5}$, and for $R_1=1$ equals the
tension of the 8 D5-\Dfive pairs in the initial model.

Let us describe the resulting spectrum, centering on the set of branes at
$z_2=0$. The 77 sector is mapped to the ${\bar 7}{\bar 7}$ sector by the
orientifold action, so we keep the former and perform no projection.
We obtain $\NN=1$ vector multiplets of $U(4)^2$, i.e. gauge bosons and
right handed fermions. The $7{\bar 7}$ sector is mapped to itself by the
orbifold and orientifold projections. The tachyonic NS groundstate
survives the (opposite) GSO projection in this sector, but the $\IZ_2$
twist projects it out. In the R sector, we obtain right-handed fermions
which, after the orientifold projection, transform in the antisymmetric
representations, $({\bf 6},{\bf 1})+({\bf 1},{\bf 6})$. The ${\ov 7}7$
spectrum is obtained analogously. Finally, the set of branes at
$z_2=(1+i)/2$ give another copy of the fields above. The complete open
string spectrum is shown in Table~\ref{tableseven}

\begin{table}[ht]
\small
\renewcommand{\arraystretch}{1.25}
\begin{center}
\begin{tabular}{|c||c|}
\hline
{\bf 77}, ${\bf{\bar 7}}{\bf{\bar 7}}$ & Gauge bosons of $U(4)^4$ \\
\hline
& Fermion$_+$ \ \ $({\bf {\rm Adj}},{\bf 1},{\bf 1},{\bf 1}) +
({\bf 1},{\bf{\rm Adj}},{\bf 1},{\bf 1}) + ({\bf 1},{\bf 1},{\bf
{\rm Adj}},{\bf 1}) + ({\bf 1},{\bf 1},{\bf 1},{\bf{\rm Adj}})$ \\
\hline
${\bf{7{\bar 7}}}$ & Fermion$_+$ \ \ $({\bf 6},{\bf 1},{\bf 1},{\bf 1}) +
({\bf 1},{\bf 6},{\bf 1},{\bf 1}) + ({\bf 1},{\bf 1},{\bf 6},{\bf 1}) +
({\bf 1},{\bf 1},{\bf 1},{\bf 6})$ \\
\hline 
${\bf{{\bar 7}7}}$ & Fermion$_+$ \ \ ({\bf 6},{\bf 1},{\bf 1},{\bf 1}) +
({\bf 1},{\bf 6},{\bf 1},{\bf 1}) + ({\bf 1},{\bf 1},{\bf 6},{\bf 1}) +
({\bf 1},{\bf 1},{\bf 1},{\bf 6}) \\
\hline 
\end{tabular}
\caption{Open string spectrum for the configuration of D7-\Dseven branes
in Figure~\ref{seven}.}
\label{tableseven}
\end{center}
\end{table}

Obviously, due to the symmetry between the two complex planes, a similar
set of transitions is obtained by varying $R_2$, while keeping $R_1$
large, so we need not repeat the analysis. However, there are new
transitions if we make both $R_1$ and $R_2$ small, which we study in what
follows. Let us keep $R_1\leq 1$, so that the initial configuration is
provided by the set of \Dtseven-branes we have just studied, and make
$R_2$ small. At $R_2=1$ the groundstate of a string stretching between
e.g. a D7-brane at $z_2=0$ and a \Dseven-brane at $z_2=(1+i)/2$ becomes
massless, and leads to a tachyon for $R_2<1$. 

The transition is very similar to those considered previously, so our
discussion is more sketchy. We may consider a configuration of 8 non-BPS
\Dteight-branes completely wrapped on the first plane and partially
wrapped on the second, and their 8 $\Omega'\alpha$ images, denoted
\Dteight$'$-branes. Such configuration would be stable only
at $R_2=1$. In order to reproduce the correct twisted charges, the
orbifold and Wilson lines Chan-Paton matrices must be
\beqa
\gamma_{\theta,{\hat 8}} = \id_8 \quad ; \quad \gamma_{W_1,{\hat 8}}
= \diag (\id_4, -\id_4) \nonumber \\
\gamma_{\theta,{\hat 8}'} = \id_8 \quad ; \quad \gamma_{W_1,{\hat 8}'}
= \diag (\id_4, -\id_4)
\eeqa
This configuration is actually T-dual to a configuration of \Dtsix-branes
identical to that considered previously (up to a relabeling of the first
and second complex planes). The open string spectrum is therefore
identical to that given in Table~\ref{tablesix}, replacing 6 and 6$'$ by 8
and 8$'$.

The stable configuration for $R_1<1$, $R_2<1$ is provided by a set of
16 D9- and 16 \Dnine-branes. It is easy to see that the total tension of
these objects agrees, for $R_1=R_2=1$, with the total tension of the 8 D5-
and 8 \Dfive-branes of the initial configuration, or, for $R_2=1$ and
arbitrary $R_1<1$, with the configuration of 8 D7-\Dseven pairs, so the
transition is energetically allowed precisely when the tachyon develops.

In order to obtain the correct twisted charges, the choice of twist and
Wilson line Chan-Paton matrices is
\beqa
\begin{array}{ccc}
\gamma_{\theta,9} = \id_{16}  \quad & \gamma_{W_1,9}=\id_{16} \quad
& \gamma_{W_2,9} = \diag(\id_8,-\id_8) \\
\gamma_{\theta,{\bar 9}}=-\id_{16} \quad & \gamma_{W_1,{\bar 9}}=
-\id_{16} \quad & \gamma_{W_2,{\bar 9}} = \diag(\id_8,-\id_8) 
\end{array}
\eeqa
This configuration is easily seen to be T-dual to a configuration with
eight D5-\Dfive pairs distributed exactly as in the configuration we
started with. Hence, the computation of the spectrum leads to
the states in Table~\ref{sixtwo}, with fivebranes replaced by ninebranes.

\begin{figure}
\begin{center}
\centering
\epsfysize=8cm
\leavevmode 
\epsfbox{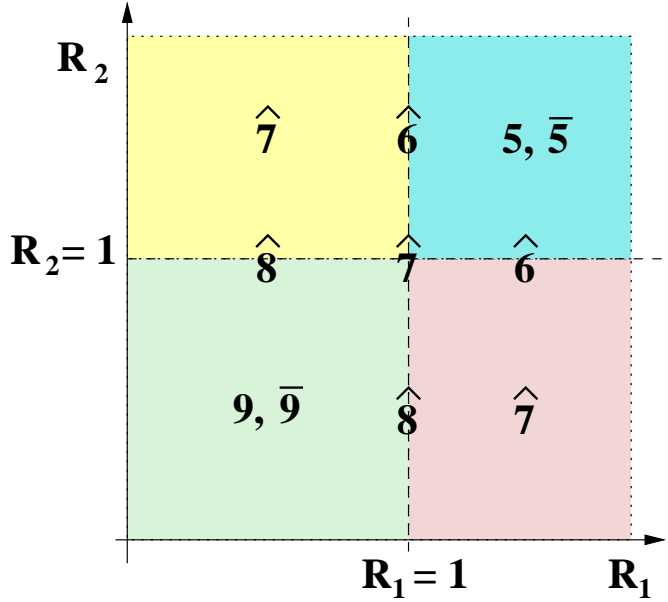}
\end{center}
\caption[]{\small Phase diagram for the $\Omega'\alpha$ orientifold of
$\IT^4/\IZ_2$. For large $R_1$, $R_2$ the stable configuration corresponds
to a set of D5- and \Dfive-branes at orbifold fixed points. When one of
the radii decreases below the critical value $R=1$, brane-antibrane pairs
decay into non-BPS \Dtseven-branes, wrapped in the small two-torus and
transverse to the large one. For $R_1,\ R_2 \ <1$, the stable configuration 
contains a set of D9- and \Dnine-branes. When one or two of the radii are
exactly equal to the critical value, there exist marginally tachyon-free
configurations. For instance, configurations of non-BPS \Dtsix-branes
partially wrapping a 1-cycle in the first plane, and transverse to the
second, for $R_1=1$, $R_2>1$; configurations of non-BPS \Dteight-branes
partially wrapping a 1-cycle in the first plane, and completely wrapped on
the second, for $R_1=1$, $R_2<1$; and a configuration of non-BPS
\Dtseven-branes wrapping one 1-cycle on each plane for $R_1=1$, $R_2=1$.}
\label{phase}  
\end{figure}

In Figure~\ref{phase} we combine the results of this section in a phase
diagram of this model, for the two-dimensional slice of parameter space
we have studied. 

\section{Conclusions}

In this paper we have studied families of type IIB orientifolds whose
closed string sector does not generate untwisted tadpoles, but generates
non-zero twisted tadpoles. In order to render the models consistent, the
required open string sector must correspond to non-supersymmetric
configurations of branes. We have discussed how to use pairs of branes and
antibranes, and non-BPS D-branes to saturate such tadpoles. The models
cannot decay to any supersymmetric vacuum, and have a rich structure of
phase transitions as the geometry of internal space is varied, which we
have explored in detail in a particular example (leaving the dynamics of
geometric moduli aside). We hope this type of analysis is useful in
understanding  the dynamics of brane-antibrane systems in more interesting
compactifications, like the semirealistic models in \cite{aiq, aiqu}.

Our analysis is related to a more general question in the orientifold
construction: Given a consistent orientifold action on the closed string
sector of a type IIB compactification, is it always possible to find a
suitable open string sectors cancelling the closed string tadpoles?
The answer is clearly `no' if one insists on introducing open string
sectors preserving the supersymmetries which are unbroken in the closed
sector. Relaxing this condition, by introducing antibranes or non-BPS
branes in the construction, leads however to cancellation of many
problematic tadpoles. For instance, untwisted tadpoles requiring a net
negative D-brane charge \cite{ads,au,aadds,kr2,kr3}, or, as explored in \cite{abg}
and in the present paper, twisted tadpoles cancelable only by D-branes for
which no untwisted tadpole exists. In this respect, we have discussed how
the $\IZ_4$ orientifold without vector structure in \cite{afiv} (and
presumably other related examples), previously thought inconsistent, can
be consistently completed by introducing brane-antibrane pairs specifically 
designed to cancel the problematic twisted tadpoles. We find these results
are encouraging and illustrate the power of brane-antibrane configurations
(and in general, non-BPS brane configurations) in cancelling closed string
tadpoles.

On the other hand, there is no general theorem ensuring this must always
be the case. Thinking in terms of the space of RR charges in a given
model, there is no guarantee that the closed string sector will define a
vector of orientifold charges for which a suitable state (brane 
configuration) in string theory exists. In fact, we have found one example
for which the kind of non-BPS configurations considered in the present
paper is not enough to achieve tadpole cancellation. The orientifold
charges lie on an empty site in the charge lattice, at least as far as the
brane configurations in the present paper are concerned. Whether more
general configurations provide states with the correct charges or not,
remains an interesting open question deserving further study.

\centerline{\bf Acknowledgements}

It is a pleasure to thank G.~Aldazabal, E.~Eyras, J.~Gomis, L.~E.~Ib\'a\~nez, 
and F.~Quevedo for useful conversations. R.~R. thanks the Theory Division
at CERN, and A.~U. thanks G.~Aldazabal and the Centro At\'omico Bariloche
(Argentina),  for hospitality during the completion of this work.
A.~U. is grateful to M.~Gonz\'alez for encouragement and support.

\appendix

\section{Miscelanea}

\subsection{Chan-Paton matrices}

In $\Omega'\alpha$ orientifolds, the general form of the Chan-Paton
matrices satisfying the algebraic consistency conditions (group law), and
defining a gauge bundle without vector structure (roughly speaking
verifying $\gamma_{\theta}^N=-\id$) are of the form 
\beqa
\gamma_{\theta} =  {\rm diag} (e^{\frac{i\pi}{N}}1_{n_1},
e^{\frac{i\pi  3}{N}}1_{n_2},\ldots, e^{\frac{i\pi (2P-1)}{N}}1_{n_{P}},
e^{\frac{i\pi (2P+1)}{N}}1_{n_{P+1}},\ldots, e^{\frac{i\pi(4P-3)}{N}}
1_{n_{N-1}}, e^{\frac{i\pi(4P-1)}{N}} 1_{n_N}) \ \ 
\label{cpone}
\eeqa
The orientifold projection $\Omega'\alpha$ implies $n_i=n_{N+1-i}$. The
structure of the orientifold actions on Chan-Paton indices is analogous
to that determined in \cite{pu}. For D5-branes (in Section 2) and
D3-branes (in Section 3) the Chan-Paton orientifold action is
{\small
\beqa
\gamma_{\Omega'\alpha} & = & {\small \pmatrix{
  &  &  &  &  & e^{\frac{i\pi}{2N}} \id_{n_1}   \cr
  &  &  &  & \cdots &    \cr
  &  &  & e^{\frac{i\pi (2P-1)}{2N}} \id_{n_P}&  &    \cr
  &  & e^{\frac{i\pi (2P+1)}{2N}} \id_{n_P} &  &  &    \cr
  & \cdots &  &  &  &    \cr
e^{\frac{i\pi (4P-1)}{2N}} \id_{n_1}  &  &  &  &  &    \cr
}}
\label{cptwosym}
\eeqa}
Since $\Omega^2=-1$ in 95 and 37$_i$ sectors \cite{gp}, the orientifold
Chan-Paton action on D9- (in Section 2) and D7$_i$-branes (in Section 3)
is given by an `antisymmetrized' version of the above matrix, namely
{\small
\beqa  
\gamma_{\Omega'\alpha} & = & {\small \pmatrix{
  &  &  &  &  & e^{\frac{i\pi}{2N}} \id_{n_1}   \cr
  &  &  &  & \cdots &    \cr
  &  &  & e^{\frac{i\pi (2P-1)}{2N}} \id_{n_P}&  &    \cr
  &  & -e^{\frac{i\pi (2P+1)}{2N}} \id_{n_P} &  &  &    \cr
  & \cdots &  &  &  &    \cr
-e^{\frac{i\pi (4P-1)}{2N}} \id_{n_1}  &  &  &  &  &    \cr
}}
\label{cptwoanti}
\eeqa}
These matrices satisfy 
\beqa
\Tr(\gamma_{\Omega'\alpha^{2k+1}}^T \gamma_{\Omega'\alpha^{}}^{-1})= \pm
\Tr \gamma_{\theta^{2k+1}}
\eeqa
with the positive and negative signs for (\ref{cptwosym}) and
(\ref{cptwoanti}), respectively.

Matrices satisfying the algebraic constraints, and defining a Chan-Paton
bundle with vector structure, are of the form
\beqa
\gamma_{\theta}= \diag(\id_{n_0}, e^{2\pi i\frac 1N} \id_{n_1},\ldots,
e^{2\pi i \frac{N-1}{N}} \id_{n_{N-1}})
\label{cpthree}
\eeqa
The orientifold projection imposes $n_{N-i}=n_i$. The $\Omega'\alpha$
Chan-Paton action for D5-branes (in section 2) and D3-branes (in Section
3) is
{\footnotesize
\beqa
\gamma_{\Omega'\alpha} = {\small \pmatrix{
\id_{v_0}  &  &  &  &  &  &  &  \cr
  &  &  &  &  &  &  & e^{2\pi i\frac 1N} \id_{n_1} \cr
  &  &  &  &  &  & \cdots &  \cr
  &  &  &  &  & e^{2\pi i\frac{P-1}{N}} \id_{n_{P-1}} &  &  \cr
  &  &  &  & e^{2\pi i \frac{P}{N}} {\bf \varepsilon}_{n_P}  &  &  &  \cr
  &  &  & e^{-2\pi i\frac{P-1}{N}} \id_{n_{P-1}} &  &  &  &  \cr
  &  & \cdots &  &  &  &  &  \cr
  & e^{-2\pi i\frac 1N} \id_{n_1} &  &  &  &  &  &  \cr
}} \quad\quad
\label{cpfoursym}
\eeqa}
For D9-branes (Section 2) and D7$_i$-branes (Section 3) the action must be
{\footnotesize
\beqa
\gamma_{\Omega'\alpha} = {\small \pmatrix{
{\bf \varepsilon}_{n_0}  &  &  &  &  &  &  &  \cr
  &  &  &  &  &  &  & e^{2\pi i\frac 1N} \id_{n_1} \cr
  &  &  &  &  &  & \cdots &  \cr
  &  &  &  &  & e^{2\pi i\frac{P-1}{N}} \id_{n_{P-1}} &  &  \cr
  &  &  &  & e^{2\pi i \frac{P}{N}} \id_{n_P}  &  &  &  \cr
  &  &  & -e^{-2\pi i\frac{P-1}{N}} \id_{n_{P-1}} &  &  &  &  \cr
  &  & \cdots &  &  &  &  &  \cr
  & -e^{-2\pi i\frac 1N} \id_{n_1} &  &  &  &  &  &  \cr
}} \quad\quad
\label{cpfouranti}
\eeqa}
They satisfy
\beqa
\Tr(\gamma_{\Omega'\alpha^{2k+1}}^T \gamma_{\Omega'\alpha^{}}^{-1})= \pm
\Tr \gamma_{\theta^{2k+1}}
\eeqa
with the positive and negative signs for (\ref{cpfoursym}) and
(\ref{cpfouranti}), respectively.

\subsection{Tadpole computations}

Tadpoles for the specific four-dimensional models studied in Section~3
have not appeared in the literature, but are easily computed following
standard techniques (see e.g. the appendix in \cite{afiv} for a general
discussion for four-dimensional orientifolds, also \cite{odriscoll,
zwart}). The only modifications are analogous to those in the six-dimensional 
case, which have been discussed in \cite{pu}. In order to keep track of
the twisted charge normalization of crosscaps with respect to disks, we
will introduce a set of D3-branes, and compute the one-loop amplitudes.
After factorization, crosscap and disk contributions can be directly read.
Disks for other types of D-branes are obtained analogously.

The twisted contributions of the cylinder, M\"obius strip and Klein bottle
in the factorization limit are
\beqa
{\cal C} & = & \sum_{k=1}^{N-1}\ {\cal C}(\theta^k)\ =\ \sum_{k=1}^{N-1}\
\prod_{i=1}^3\ (2\sin(\pi k v_i)\;)\ (\Tr \gamma_{\theta^k,3})^2
\nonumber \\ 
{\cal M} & = & {\cal M} (\Omega'\alpha^{2k+1})\ =\ \sum_{k=0}^{N-1}\
8\ \times \prod_{i=1}^3\ 2\sin (\pi(2k+1)v_i/2)\
\Tr(\,\gamma_{\Omega'\alpha^{2k+1}}^T\, 
\gamma_{\Omega'\alpha^{2k+1}}^{-1}\,) \nonumber \\
{\cal K} & = & \sum_{k=0}^{N-1} [\ {\cal K} (1,\Omega'\alpha^{2k+1}) +
{\cal K} (\theta^{N/2}, \Omega' \alpha^{2k+1}) \ ]  =  \nonumber \\
& = & \sum_{k=0}^{N-1} 16 \left[ \; \prod_{i=1}^3\
\frac{2\sin(\pi(2k+1)v_i)}{4\cos^2(\pi(2k+1)v_i/2)}\ \pm\ 
\frac{2\sin(\pi(2k+1)v_3)}{4\cos^2(\pi(2k+1)v_3/2)} \; \right]
\eeqa
with the sign in the Klein bottle is positive (negative) for the {\bf A}-
type ({\bf B}-type) projection.
 
The cylinder can be recast as
\beqa
{\cal C} & = & \sum_{k=1}^{N-1}\ \frac{1}{\prod_{i=1}^3 2\sin(\pi k v_i)}
\ [\ \prod_{i=1}^3 [2 \sin(\pi k v_i)] \Tr \gamma_{\theta^k,3} \ ]\; ^2
\eeqa
The matrices in section A.1 satisfy
\beqa
\Tr (\, \gamma_{\Omega'\alpha^{2k+1},3}^T\
\gamma_{\Omega'\alpha^{2k+1},3}^{-1}\,)\ = 
\pm\ \Tr (\, \gamma_{\Omega'\alpha^{2k+1+N},3}^T\ 
\gamma_{\Omega'\alpha^{2k+1+N},3}^{-1}\,)\ = \Tr \gamma_{\theta^{2k+1},3}
\eeqa
with the upper (lower) sign for matrices with (without) vector structure,
which for D3-branes is correlated with the {\bf B}-type ({\bf A}-type)
projection. This allows to rewrite the Moebius strip contribution as
\beqa
{\cal M} \ = \ 
\sum_{k=0}^{N/2-1} \frac{1}{8\, s_1 s_2 s_3} \;
2\times [\; 8 s_1 s_2 s_3\, \Tr\gamma_{\theta^{2k+1},3}\; ]\ 
[\ 32\ (\; \ts_1 \ts_2 \ts_3 \pm \tc_1 \tc_2 \ts_3 \; ) \ ]
\eeqa
with the upper (lower) for the {\bf A}-type ({\bf B}-type) projection
(notice the additional sign flip in shifting $k\to k+N/2$ in the
trigonometric factors). Also we have introduced 
\beqa
s_i & = \sin(\pi (2k+1)v_i),\quad \ts_i &= \sin(\pi (2k+1)v_i/2),
\nonumber\\
c_i  &= \cos(\pi (2k+1)v_i), \quad \tc_i & = \cos(\pi (2k+1)v_i/2)
\eeqa

Finally, the Klein bottle can be rewritten as
\beqa
{\cal K} & = & 
\sum_{k=0}^{N/2-1}  16\; {{1}\over {8\, s_1 s_2 s_3}} \;
[\ \frac{8\, s_1 s_2 s_3 \ts_1 \ts_2 \ts_3}{\tc_1 \tc_2 \tc_3}\;
+\; \frac{8\, s_1 s_2 s_3 \tc_1 \tc_2 \ts_3}{\ts_1 \ts_2 \tc_3}\;
\pm\; \frac{16\, s_1 s_2 s_3 \ts_3}{\tc_3} \; ]\ = \nonumber \\
 &= & \sum_{k=0}^{N/2-1}  \frac{1}{ 8\, s_1 s_2 s_3} \; 
[\; 32 (\, \ts_1\, \ts_2\, \ts_3\, \pm\, \tc_1\, \tc_2\, \tc_3 \,) \;]\;^2
\eeqa
with the upper (lower) sign for the {\bf A}-type ({\bf B}-type)
projection. 

The total contribution ${\cal C}+{\cal M}+{\cal K}$ factorizes as
\beqa
& \sum_{k=0}^{N-1} {{1}\over{\prod_{i=1}^3 2\sin(\pi k v_i )}}   
\left[\, \prod_{i=1}^3 2 \sin(\pi k v_i)\, \Tr \gamma_{\theta^k,3} \right.
& +\\
& + 32\, \delta_{k,1\ {\rm mod}\ 2} \sin(\pi k v_3/2)
( \sin(\pi k v_1/2) \sin(\pi k v_2/2)  & \left. \pm  
\cos(\pi k v_1/2) \cos(\pi k v_2/2)\, ) \,\right]^2  \nonumber
\eeqa
from which, after use of trigonometric identities, we can read off the
crosscap tadpoles 
\beqa
\begin{array}{lll}
{\cal T}_k & =  -16  \delta_{k,1\; {\rm mod}\; 2}\ 
(\ \sin(\pi k v_1) + \sin(\pi k v_2) \ ) & {\rm for}
\; {\rm the}\; {\rm {\bf A}-type} \; {\rm projection} \\
{\cal T}_k & =  +16  \delta_{k,1\; {\rm mod}\; 2}\ \sin(\pi k v_3) &
 {\rm for} \; {\rm the}\; {\rm {\bf B}-type} \; {\rm projection}
\end{array}
\eeqa
in units in which the D3-brane disk tadpole is 
$\prod_i 2\sin(\pi kv_i) \Tr \gamma_{\theta^k,3}$. It is straightforward
to check (for instance, from $37_i$ cylinders) that in these units the
twisted charges of a D7$_i$-brane is $2\sin(\pi k v_i)
\Tr\gamma_{\theta^k,7_i}$.

\end{document}